\documentclass[runningheads]{llncs}

\usepackage[utf8]{inputenc}
\usepackage{amsfonts,amssymb,latexsym}
\usepackage{amsmath,amssymb,braket,tikz}
\usepackage{upgreek}
\usepackage{xcolor}
\usepackage{epic}
\usepackage{cite}
\usepackage{eucal}
\usepackage{graphicx}
\usetikzlibrary{tikzmark,calc}
\usetikzlibrary{arrows.meta,decorations.pathreplacing,calligraphy}

\usepackage{algorithm}
\usepackage[noend]{algorithmic}

\usepackage{changes}

\colorlet{mylightblue}{cyan!30}
\colorlet{mylightgray}{lightgray!50}
\colorlet{mypink}{red!30}
\colorlet{mylime}{lime!70}
\colorlet{lightyellow}{yellow!50}
\colorlet{lightpurple}{blue!20}

\colorlet{telD}{mylightgray}
\colorlet{telS}{mylightblue}
\colorlet{edgeC}{cyan}
\colorlet{edgeM}{gray}
\colorlet{edgeD}{black} 
\colorlet{edgeS}{red}
\colorlet{firstEdge}{blue}

\colorlet{path}{brown!90!blue} 

\definecolor{darkestblue}{rgb}{0,0,0.4}
\colorlet{darkblue}{blue!70!black}
\colorlet{darkred}{red!70!black}
\colorlet{fam1}{red!70!black}
\colorlet{fam3}{blue!70}
\colorlet{fam2}{green!70!black}
\colorlet{fam5}{orange!60}
\colorlet{fam4}{magenta!60}

\newcommand{\IFTHEN}[1]{
  \STATE \algorithmicif\ #1\ \algorithmicthen}

\renewcommand\Tilde[1]{
  \widetilde{#1}
}
\newcommand\TTilde[1]{
  \setlength{\arraycolsep}{0em}\begin{array}[b]{c}\begin{array}[b]{c}\widetilde{}\\[-2.3ex]\widetilde{}\end{array}\\[-2.3ex]#1\end{array}
}

\newcommand\B[1]{\mathsf{#1}}
\newcommand\T[1]{{\tt #1}}

\newcommand\g[1]{\mathbb{#1}}

\newcommand\z{\mspace{2mu}}

\newcommand\rev[1]{\overline{#1}}

\newcommand\ppl{\{2,\!4\}}
\newcommand\cpl{\{4,\!6\}}

\newcommand\nodecan[3]{\begin{picture}(0,0)
		\thicklines
	    \put(0,0){\color{#3}\circle*{14}}
        \put(0,0){\color{black}\circle{14}}
        \put(0,0){\makebox(0,0){$\mathtt{#1}^{#2}$}}
\end{picture}} 

\newcommand\nodeidx[4]{\begin{picture}(0,0)
		\thicklines
		\put(0,0){\color{#4}\circle*{16}}
		\put(0,0){\color{black}\circle{16}}
		
		\put(0,0){\makebox(0,0){$\T{#1}^{#2}_\T{#3}$}}
\end{picture}}

\newcommand\mysquareidA[9]{\begin{picture}(0,0)
     {\linethickness{1.5pt}
		\put(0,30){\color{red}\line(1,0){30}}
		\put(0,0){\color{red}\line(1,0){30}}
		\put(0,0){\color{red}\line(0,1){30}}
		\put(30,0){\color{red}\line(0,1){30}}
	}

    \thicklines
		\put(0,30){\nodeidx{#1}{#2}{#5}{white}}
 
        \put(30,0){\nodeidx{#1}{#2}{#8}{mylightgray}}
        
        \put(30,30){\nodeidx{#3}{#4}{#6}{white}}
        
        \put(0,0){\nodeidx{#3}{#4}{#7}{white}}
        
        \put(15,15){\makebox(0,0){\bf\color{darkred}#9}}
\end{picture}} 

\newcommand\mysquareidB[9]{\begin{picture}(0,0)
     {\linethickness{1.5pt}
		\put(0,30){\color{red}\line(1,0){30}}
		\put(0,0){\color{red}\line(1,0){30}}
		\put(0,0){\color{red}\line(0,1){30}}
		\put(30,0){\color{red}\line(0,1){30}}
	}

    \thicklines
		\put(0,30){\nodeidx{#1}{#2}{#5}{white}}
 
        \put(30,0){\nodeidx{#1}{#2}{#8}{white}}
        
        \put(30,30){\nodeidx{#3}{#4}{#6}{white}}
        
        \put(0,0){\nodeidx{#3}{#4}{#7}{mylightgray}}
        
        \put(15,15){\makebox(0,0){\bf\color{darkred}#9}}
\end{picture}}

\newcommand\mysquaretelomeres[8]{\begin{picture}(0,0)
    \thicklines
		\put(0,30){\nodeidx{#1}{#2}{#5}{mylightblue}}
 
        \put(30,0){\nodeidx{#1}{#2}{#8}{lightpurple}}
        
        \put(30,30){\nodeidx{#3}{#4}{#6}{lightpurple}}
        
        \put(0,0){\nodeidx{#3}{#4}{#7}{mylightblue}}
        
\end{picture}}

\newcommand\mysquarehA[9]{\begin{picture}(0,0)
		{\linethickness{1.5pt}
		\put(0,30){\color{cyan}\line(1,0){30}}
		\put(0,0){\color{cyan}\line(1,0){30}}
		
		\put(0,0){\color{lightgray}\thinlines\line(0,1){30}}
		\put(30,0){\color{lightgray}\thinlines\line(0,1){30}}
	}

    \thicklines
		\put(0,30){\nodeidx{#1}{#2}{#5}{white}}
 
        \put(30,0){\nodeidx{#1}{#2}{#8}{mylightgray}}
        
        \put(30,30){\nodeidx{#3}{#4}{#6}{white}}
        
        \put(0,0){\nodeidx{#3}{#4}{#7}{white}}
        
        \put(15,15){\makebox(0,0){\color{gray} #9}}
\end{picture}}

\newcommand\mysquarehB[9]{\begin{picture}(0,0)
		{\linethickness{1.5pt}
		\put(0,30){\color{cyan}\line(1,0){30}}
		\put(0,0){\color{cyan}\line(1,0){30}}
		
		\put(0,0){\color{lightgray}\thinlines\line(0,1){30}}
		\put(30,0){\color{lightgray}\thinlines\line(0,1){30}}
	}

    \thicklines
		\put(0,30){\nodeidx{#1}{#2}{#5}{white}}
 
        \put(30,0){\nodeidx{#1}{#2}{#8}{white}}
        
        \put(30,30){\nodeidx{#3}{#4}{#6}{white}}
        
        \put(0,0){\nodeidx{#3}{#4}{#7}{mylightgray}}
        
        \put(15,15){\makebox(0,0){\color{gray} #9}}
\end{picture}}

\newcommand{\pv}[3]{\node[shape=circle,draw=black, fill=white, scale=1.5] (#2) at #1 {};
\node at #1 {#3};	
}

\newcommand{\pvcl}[4]{\node[shape=circle,draw=black, fill=#4, scale=1.6] (#2) at #1 {};
\node at #1 {#3};	
}

\newcommand{\vgr}[3]{
\node[shape=circle,draw=lightgray,thick, fill=white, scale=0.6] (#2) at #1 {};
\node at #1 {\footnotesize #3};	
}

\newcommand{\vcl}[3]{
\node[shape=circle,draw=black,thick, fill=white, scale=0.6] (#2) at #1 {};
\node at #1 {\footnotesize #3};	
}

\newcommand{\tvcl}[4]{
\node[shape=circle,draw=black,thick, fill=#4, scale=0.9] (#2) at #1 {};
\node at #1 {\scriptsize #3};	
}

\newcommand{\vsm}[3]{\node[shape=circle,draw=#3,thick, fill=white, scale=0.6] (#2) at #1 {};
}

\newcommand{\cvgr}[3]{\node[shape=circle,draw=gray, fill=white, scale=2.5] (#2) at #1 {};
\node at #1 {\color{gray}#3};	
}

\newcommand{\cvcl}[4]{\node[shape=circle,draw=black, fill=#4, scale=2.5] (#2) at #1 {};
\node at #1 {#3};	
}

\newcommand{\dted}[3]{\path [dotted, ultra thick, draw=#3] (#1) edge (#2);}

\newcommand{\dtedthin}[3]{\path [dotted, draw=#3] (#1) edge (#2);}

\newcommand{\ded}[3]{\path [dashed, thick, draw=#3] (#1) edge (#2);}
\newcommand{\ed}[3]{\path [ultra thick, draw=#3] (#1) edge (#2);}
\newcommand{\edthin}[3]{\path [thick, draw=#3] (#1) edge (#2);}

\usepackage{geometry}

\newgeometry{textwidth=15cm,textheight=21cm}
\newgeometry{left=2cm, right=2cm, top=2.3cm, bottom=2.3cm, footskip=8mm}

\title{Investigating the complexity of the double distance problems}

\author{Mar\'ilia D. V. Braga
\and
Leonie R. Brockmann
\and
Katharina Klerx
\and
Jens Stoye
}

\institute{Faculty of Technology and CeBiTec, 
Bielefeld University, 
Germany
}

\authorrunning{M.\,D.\,V. Braga {\it et al.}} 

\begin{document}

\maketitle

\begin{abstract} 
Two genomes $\g{A}$ and $\g{B}$ over the same set of gene families form a \emph{canonical} pair when each of them has exactly one gene from each family.
Different distances of canonical genomes can be derived from a structure called \emph{breakpoint graph}, which represents the relation between the two given genomes as a collection of cycles of even length and paths. Let $c_i$ and $p_j$ be the numbers of cycles of length $i$ and of paths of length $j$, respectively. Furthermore, let $n_*$ be the number of common genes of genomes $\g{A}$ and $\g{B}$.
Then, the breakpoint distance of $\g{A}$ and $\g{B}$ is equal to $n_*-\left(c_2+\frac{p_0}{2}\right)$. Similarly, when the considered rearrangements are those modeled by the \emph{double-cut-and-join} (DCJ) operation, the rearrangement distance of $\g{A}$ and $\g{B}$ is $n_*-\left(c+\frac{p_\textup{e}}{2}\right)$, where $c$ is the total number of cycles and $p_\textup{e}$ is the total number of paths of even length.

\smallskip
The distance formulation is a basic unit for several other combinatorial problems related to genome evolution and ancestral reconstruction, such as \emph{median} or \emph{double distance}.
Interestingly, both median and double distance problems can be solved in polynomial time for the breakpoint distance, while they are NP-hard for the rearrangement distance.
One way of exploring the complexity space between these two extremes is to consider a $\sigma_k$ distance, defined to be $n_*-\left(c_2+c_4+\ldots+c_k+\frac{p_0+p_2+\ldots+p_{k-2}}{2}\right)$, and increasingly investigate the complexities of median and double distance for the $\sigma_4$ distance, then the~$\sigma_6$ distance, and so on.
While for the median much effort was done in our and in other research groups but no progress was obtained even for the $\sigma_4$ distance, for solving the double distance under $\sigma_4$ and $\sigma_6$ distances we could devise linear time algorithms, which we present here.

\keywords{Comparative genomics, Genome rearrangement, breakpoint distance, double-cut-and-join (DCJ) distance, double distance.}
\end{abstract}




\section{Introduction}

In genome comparison, the most elementary problem is that of computing a \emph{distance} between two given genomes~\cite{San92}, each one being a set of \emph{chromosomes}. Usually a high-level view of a chromosome is adopted, in which each chromosome is represented by a sequence of oriented \emph{genes} and the genes are classified into \emph{families}. 
The simplest model in this setting is the \emph{breakpoint} model, whose distance consists of somehow quantifying the distinct \emph{adjacencies} between the two genomes, an \emph{adjacency} in a genome being the oriented neighborhood between two genes in one of its chromosomes~\cite{TAN-ZHE-SAN-2009}. Other models rely on large-scale genome \emph{rearrangements}, such as inversions, translocations, fusions and fissions, yielding distances that correspond to the minimum number of rearrangements required to transform one genome into another~\cite{HAN-PEV-1995,HAN-PEV-1999,YAN-ATT-FRI-2005}. 

Independently of the underlying model, the distance formulation is a basic unit for several other combinatorial problems related to genome evolution and ancestral reconstruction~\cite{TAN-ZHE-SAN-2009}. The \emph{median} problem, for example, has three genomes as input and asks for an ancestor genome that minimizes the sum of its distances to the three given genomes. Other models are related to the \emph{whole genome duplication} (WGD) event~\cite{ELM-SAN-2003}.
Let the \emph{doubling} of a genome duplicate each of its chromosomes. 
The \emph{double distance} is the problem that has a \emph{duplicated} genome and a \emph{singular} genome as input and computes the distance between the former and a doubling of the latter. The \emph{halving} problem has a duplicated genome as input and asks for a singular genome whose double distance to the given duplicated genome is minimized. Finally, the \emph{guided halving} problem has a duplicated and a singular genome as input and asks for another singular genome that minimizes the sum of its double distance to the given duplicated genome and its distance to the given singular genome.

Our study relies on 
the \emph{breakpoint graph}, a structure that represents the relation between two given genomes~\cite{BAF-PEV-1993}. When the two genomes are over the same set of gene families and form a \emph{canonical} pair, that is, when each of them has exactly one gene from each family, their breakpoint graph is a collection of cycles of even length and paths. Assuming that both genomes have $n_*$ genes, if we call \emph{$k$-cycle} a cycle of length~$k$ and \emph{$k$-path} a path of length~$k$, the corresponding breakpoint distance is equal to~$n_*-\left(c_2+\frac{p_0}{2}\right)$, where $c_2$ is the number of 2-cycles and $p_0$ is the number of 0-paths~\cite{TAN-ZHE-SAN-2009}. Similarly, when the considered rearrangements are those modeled by the \emph{double-cut-and-join} (DCJ) operation~\cite{YAN-ATT-FRI-2005}, the rearrangement distance is $n_*-\left(c+\frac{p_e}{2}\right)$, where $c$ is the total number of cycles and $p_e$ is the total number of even paths~\cite{BER-MIX-STO-2006}. 

While the halving problem under both breakpoint and rearrangement distances can be solved in polynomial time~\cite{TAN-ZHE-SAN-2009,ELM-SAN-2003,ALE-PEV-2008,MIX-2008}, median, double distance and guided halving problems can be solved in polynomial time only under the breakpoint distance, but are NP-hard under the rearrangement distance~\cite{TAN-ZHE-SAN-2009}.
One way of exploring the complexity space between these two extremes is to consider a~$\sigma_k$ distance~\cite{Chauve2018}, defined to be $n_*-\left(c_2+c_4+\ldots+c_k+\frac{p_0+p_2+\ldots+p_{k-2}}{2}\right)$, and increasingly investigate the complexities of median, guided halving and double distance under the $\sigma_4$ distance, then under the $\sigma_6$ distance, and so on. Note that the $\sigma_2$ distance is the breakpoint distance and the $\sigma_\infty$ distance is the DCJ distance. To the best of our knowledge, the guided halving problem has not been studied for this class of problems, while for the median under $\sigma_4$ distance much effort has been done in our group and in other research groups (e.g.~\cite{Chauve2018}) but no progress was obtained so far.

In contrast, for the double distance, while~$\sigma_8$ and higher were not yet studied, we succeeded in devising efficient algorithms for $\sigma_4$ and $\sigma_6$. Our results, which we present here, are built on a variation of the breakpoint graph, called \emph{ambiguous breakpoint graph}~\cite{TAN-ZHE-SAN-2009} and have three main parts.
First we show that in any $\sigma_k$ double distance, including the NP-hard DCJ double distance, all 2-cycles and 0-paths are fulfilled, meaning that the common adjacencies and common telomeres between the compared genomes are always conserved.
Then we show that the  $\sigma_4$ double distance can be computed by a greedy linear time algorithm.
Finally we present a non-greedy but still linear time algorithm for the $\sigma_6$ double distance. 

This paper is an extended version of our two recent works~\cite{BBKS2022,BBKS2023}.

\section{Background}

\begin{figure}[ht!]

  \begin{center}
  \includegraphics[scale=0.36]{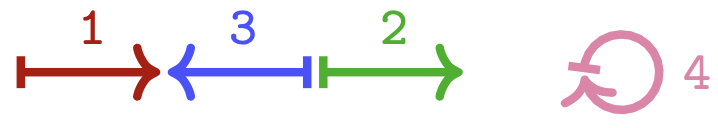}
  \end{center}
  
  \caption{Representation of linear chromosome $[\T{1}\z\rev{\T{3}}\z\T{2}]$ and circular chromosome~$(\T{4})$.}\label{fig:chr}
  \end{figure}

A \emph{chromosome} is an oriented DNA molecule and can be either linear or circular. We represent a chromosome by its sequence of genes, where each \emph{gene} is an oriented DNA fragment. We assume that each gene belongs to a \emph{family}, which is a set of homologous genes. 
A gene that belongs to a family $\mathtt{X}$ is represented by the symbol $\mathtt{X}$ itself if it is read in forward orientation or by the symbol $\overline{\mathtt{X}}$ if it is read in reverse orientation.
For example, the sequences $[\mathtt{1}\,\overline{\mathtt{3}}\,\mathtt{2}]$ and $(4)$  represent, respectively, a linear (flanked by square brackets) and a circular chromosome (flanked by parentheses), both shown in Figure~\ref{fig:chr}, the first composed of three genes 
and the second composed of a single gene.
Note that if a sequence $s$ represents a chromosome $K$, then $K$ can be equally represented by the \emph{reverse complement} of~$s$, denoted by $\overline{s}$, obtained by reversing the order and the orientation of the genes in $s$. Moreover, if $K$ is circular, it can be equally represented by any circular rotation of $s$ and  $\overline{s}$. Recall that a gene is an \emph{occurrence} of a family, therefore distinct genes from the same family are  represented by the same symbol.

We can also represent a gene from family $\T{X}$ referring to its \emph{extremities} $\T{X}^h$ (head) and $\T{X}^t$ (tail).
The \emph{adjacencies} in a chromosome are the neighboring extremities of distinct genes. The remaining extremities, that are at the ends of linear chromosomes, are \emph{telomeres}.
In linear chromosome $[\T{1}\z\rev{\T{3}}\z\T{2}]$, the adjacencies are $\{\T{1}^h\T{3}^h, \T{3}^t\T{2}^t\}$ and the telomeres are $\{\T{1}^t,\T{2}^h\}$. 
Note that an adjacency has no orientation, that is, an adjacency between extremities $\T{1}^h$ and $\T{3}^h$ can be equally represented by $\T{1}^h\T{3}^h$ and by $\T{3}^h\T{1}^h$.
In the particular case of a single-gene circular chromosome, e.g.~$(\T{4})$, an adjacency exceptionally occurs between the extremities of the same gene (here $\T{4}^h\T{4}^t$).

A \emph{genome} is then a multiset of chromosomes and we denote by~$\mathcal{F}(\mathbb{G})$ the set of gene families that occur in genome $\mathbb{G}$.
In addition, we denote by~$\mathcal{A}(\mathbb{G})$ the multiset of adjacencies and by~$\mathcal{T}(\mathbb{G})$ the multiset of telomeres that occur in~$\mathbb{G}$.
A genome $\mathbb{S}$ is called \emph{singular} if each gene family occurs exactly once in $\mathbb{S}$.
Similarly, a genome $\mathbb{D}$ is called \emph{duplicated} if each gene family occurs exactly twice in $\mathbb{D}$. The two occurrences of a family in a duplicated genome are called \emph{paralogs}. 
A \emph{doubled} genome is a special type of duplicated genome in which each adjacency or telomere occurs exactly twice.
These two copies of the same adjacency (respectively same telomere) in a doubled genome are called \emph{paralogous adjacencies} (respectively \emph{paralogous telomeres}).
Observe that distinct doubled genomes with circular chromosomes can have exactly the same adjacencies and telomeres, as we show in Table~\ref{tab:genome-types}, where we also give examples of singular and duplicated genomes.

\begin{table}[ht!]
 \caption{\label{tab:genome-types}
  Examples of a singular, a duplicated and two doubled genomes, with their sets of families and their multisets of adjacencies. Note that the doubled genomes $\mathbb{B}_1$ and $\mathbb{B}_2$ have exactly the same adjacencies and telomeres.}
  
  \centering
  \scriptsize
  \setlength{\tabcolsep}{1pt}
  
  \hspace{-3mm}
  \begin{tabular}{cll}
    \hline
   & &  \\[-2ex]    
   \parbox{3.6cm}{Singular genome\\(each family occurs once)} & $\mathbb{S}\!=\!\{(\T{1}\z\rev{\T{3}}\z\T{2})\z(\T{4})[\T{5}\z\rev{\T{6}}]\}$ & \parbox{4.8cm}{$\begin{cases}\mathcal{F}(\mathbb{S})\!=\!\{\T{1},\T{2},\T{3},\T{4},\T{5},\T{6}\}\\\mathcal{A}(\mathbb{S})\!=\!\{\mathtt{1}^h\mathtt{3}^h, \mathtt{3}^t\mathtt{2}^t, \mathtt{2}^h\mathtt{1}^t, \mathtt{4}^h\mathtt{4}^t, \mathtt{5}^h\mathtt{6}^h\}\\\mathcal{T}(\mathbb{S})\!=\!\{\mathtt{5}^t, \mathtt{6}^t\}\end{cases}$}\\[4ex]
    \hline
    & &  \\[-2ex] 
    \parbox{3.6cm}{Duplicated genome\\(each family occurs twice)} & $\mathbb{D}\!=\!\{(\mathtt{1\,2}\,\overline{\mathtt{3}}\,\mathtt{1})[\overline{\mathtt{3}}\,\mathtt{2}]\}$& \parbox{4.8cm}{$\begin{cases}\mathcal{F}(\mathbb{D})\!=\!\{\T{1},\T{2},\T{3}\}\\
    \mathcal{A}(\mathbb{D})\!=\!\{\mathtt{1}^h\mathtt{2}^t, \mathtt{2}^h\mathtt{3}^h, \mathtt{3}^t\mathtt{1}^t, \mathtt{1}^h\mathtt{1}^t, \mathtt{3}^t\mathtt{2}^t\}\\\mathcal{T}(\mathbb{D})\!=\!\{\mathtt{3}^h, \mathtt{2}^h\}\end{cases}$}\\[4ex]
    \hline
    & &  \\[-2ex]  
    \parbox{3.6cm}{Doubled genomes\\(each adj.\,or tel.\,occurs twice)} & \parbox{3.15cm}{$\mathbb{B}_1\!=\!\{(\T{1}\,\T{2})\,(\T{1}\,\T{2})\,[\T{3}\,\T{4}]\,[\T{3}\,\T{4}]\}$\\$\mathbb{B}_2\!=\!\{(\T{1}\,\T{2}\,\T{1}\,\T{2})\,[\T{3}\,\T{4}]\,[\T{3}\,\T{4}]\}$} &\parbox{4.8cm}{$\begin{cases}\mathcal{F}(\mathbb{B}_i)\!=\!
    \{\T{1},\T{2},\T{3},\T{4}\}\\
    \mathcal{A}(\mathbb{B}_i)\!=\!\{\mathtt{1}^h\!\mathtt{2}^t\!, \mathtt{2}^h\!\mathtt{1}^t\!, \mathtt{1}^h\!\mathtt{2}^t\!, \mathtt{2}^h\!\mathtt{1}^t\!, \mathtt{3}^h\!\mathtt{4}^t\!, \mathtt{3}^h\!\mathtt{4}^t\}\\\mathcal{T}(\mathbb{B}_i)\!=\!\{\mathtt{3}^t,\mathtt{4}^h,\mathtt{3}^t, \mathtt{4}^h\}\end{cases}$}\\[4ex]
    \hline
  \end{tabular}
\end{table}

\subsection{Comparing canonical genomes}
Two  genomes $\mathbb{S}_1$ and $\mathbb{S}_2$ are said to be a \emph{canonical pair} when they are singular and have the same gene families, that is, $\mathcal{F}(\mathbb{S}_1)=\mathcal{F}(\mathbb{S}_2)$.
Denote by~$\mathcal{F}_*$ the set of families occurring in canonical genomes $\mathbb{S}_1$ and $\mathbb{S}_2$, and by $n_* = |\mathcal{F}_*|$ its cardinality.
For example, genomes $\mathbb{S}_1=\{(\T{1}\z\rev{\T{3}}\z\T{2})\z(\T{4})\}$ and $\mathbb{S}_2=\{(\T{1}\z\T{2})\z(\T{3}\z\rev{\T{4}})\}$ are canonical with $\mathcal{F}_*=\{\T{1},\T{2},\T{3},\T{4}\}$ and $n_* = 4$.

\subsubsection{Breakpoint graph.}

The relation between two canonical genomes $\mathbb{S}_1$ and $\mathbb{S}_2$ can be represented by their \emph{breakpoint graph} $BG(\mathbb{S}_1, \mathbb{S}_2) = (V,E)$, that is a multigraph representing the adjacencies of $\mathbb{S}_1$ and $\mathbb{S}_2$~\cite{BAF-PEV-1993}.
The vertex set~$V$ comprises, for each family $\mathtt{X}$ in $\mathcal{F}_*$, one vertex for the extremity $\mathtt{X}^h$ and one vertex for the extremity $\mathtt{X}^t$.
The edge multiset $E$ represents the adjacencies.
For each adjacency in $\mathbb{S}_1$ there exists one $\mathbb{S}_1$-edge in $E$ linking its two extremities. Similarly, for each adjacency in $\mathbb{S}_2$ there exists one $\mathbb{S}_2$-edge in $E$ linking its two extremities.
Clearly, $BG(\mathbb{S}_1, \mathbb{S}_2)$ can easily be constructed in linear $O(n_*)$ time.

\begin{figure}[ht!]
  \begin{center}
  \includegraphics[scale=0.36]{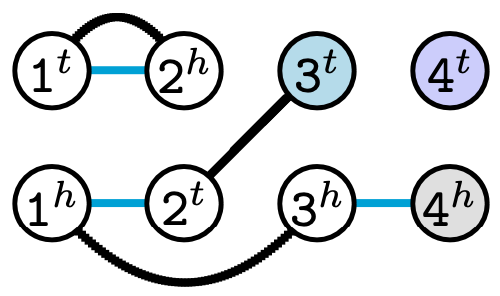}
  
  \end{center}
  
  \caption{\label{fig:bg}Breakpoint graph of genomes
  $\mathbb{S}_1=\{\,(\mathtt{1\,2})\,[\mathtt{3}\,\overline{\mathtt{4}}]\,\}$ and $\mathbb{S}_2=\{\,(\mathtt{1}\,\overline{\mathtt{3}}\,\mathtt{2})\,[\mathtt{4}]\,\}$. Edge types are distinguished by colors: $\mathbb{S}_1$-edges are drawn in blue and $\mathbb{S}_2$-edges are drawn in black. Similarly, vertex types are distinguished by colors: an $\g{S}_1$-telomere is marked in blue, an $\g{S}_2$-telomere is marked in gray, a telomere in both $\g{S}_1$ and $\g{S}_2$ is marked in purple and non-telomeric vertices are white. This graph has one 2-cycle, one 0-path and one 4-path.}
\end{figure}

The degree of each vertex can be 0, 1 or 2 and each connected \emph{component} alternates between $\mathbb{S}_1$- and $\mathbb{S}_2$-edges. As a consequence, the components of the breakpoint graph of canonical genomes can be cycles of even length or paths. An even path has one endpoint in $\g{S}_1$ (\emph{$\g{S}_1$-telomere}) and the other in $\g{S}_2$ (\emph{$\g{S}_2$-telomere}), while an odd path has either both endpoints in $\g{S}_1$ or both endpoints in $\g{S}_2$. 
A vertex that is not a telomere in $\g{S}_1$ nor in $\g{S}_2$ is said to be \emph{non-telomeric}. In the breakpoint graph a non-telomeric vertex has degree 2.
We call \emph{$i$-cycle} a cycle of length~$i$ and \emph{$j$-path} a path of length~$j$. We also denote by $c_i$ the number of $i$-cycles, by $p_j$ the number of $j$-paths, by $c$ the total number of cycles and by $p_\textup{e}$ the total number of even paths. Since the number of telomeres in each genome is even (2 telomeres per linear chromosome), the total number of even paths in the breakpoint graph must be even.
An example of a breakpoint graph is given in Figure~\ref{fig:bg}.

\subsubsection{Breakpoint distance.}

For canonical genomes $\mathbb{S}_1$ and $\mathbb{S}_2$ the \emph{breakpoint distance}, denoted by $\textup{d}_\textsc{bp}$, is defined as follows~\cite{TAN-ZHE-SAN-2009}:

$$\textup{d}_\textsc{bp}(\mathbb{S}_1, \mathbb{S}_2)=n_*-\left(|\mathcal{A}(\mathbb{S}_1)\cap\mathcal{A}(\mathbb{S}_2)|+\frac{|\mathcal{T}(\mathbb{S}_1)\cap\mathcal{T}(\mathbb{S}_2)|}{2}\right).$$

For $\mathbb{S}_1=\{(\T{1}\z\rev{\T{3}}\z\T{2})\z[\T{4}]\}$ and $\mathbb{S}_2=\{(\T{1}\z\T{2})\z[\T{3}\z\rev{\T{4}}]\}$, we have $n_*=4$. The set of common adjacencies is $\mathcal{A}(\mathbb{S}_1)\cap\mathcal{A}(\mathbb{S}_2)=\{\T{1}^t\T{2}^h\}$ and the set of common telomeres is $\mathcal{T}(\mathbb{S}_1)\cap\mathcal{T}(\mathbb{S}_2)=\{\T{4}^t\}$, giving $\textup{d}_\textsc{bp}(\mathbb{S}_1, \mathbb{S}_2)=2.5$.
Since a common adjacency of $\mathbb{S}_1$ and $\mathbb{S}_2$ corresponds to a 2-cycle and a common telomere corresponds to a 0-path in $BG(\mathbb{S}_1, \mathbb{S}_2)$, the breakpoint distance can be rewritten as

$$\textup{d}_\textsc{bp}(\mathbb{S}_1, \mathbb{S}_2)=n_*-\left(c_2+\frac{p_0}{2}\right).$$

\subsubsection{DCJ distance.}

Given a genome, a \emph{double cut and join} (DCJ) is the operation that breaks two of its adjacencies or telomeres\footnote{A broken adjacency has two open ends and a broken telomere has a single one.} and rejoins the open extremities in a different way~\cite{YAN-ATT-FRI-2005}. 
For example, consider the chromosome $K=[\z\T{1}\z\T{2}\z\T{3}\z\T{4}\z]$ and a DCJ that cuts $K$ between genes~$\T{1}$ and~$\T{2}$ and between genes~$\T{3}$ and~$\T{4}$, creating segments $\T{1}\bullet$, $\bullet\T{2}\z\T{3}\bullet$ and $\bullet\T{4}$ (where the symbols $\bullet$ represent the open ends). If we join the first with the third and the second with the fourth open end, we get~$K'=[\z\T{1}\z\rev{\T{3}}\z\rev{\T{2}}\z\T{4}\z]$, that is, the described DCJ operation is an inversion transforming~$K$ into~$K'$.
Besides inversions, DCJ operations can represent several rearrangements, such as translocations, fissions and fusions.
The \emph{DCJ distance} $\textup{d}_\textsc{dcj}$ is then the minimum number of DCJs that transform one genome into the other and can be easily computed with the help of their breakpoint graph~\cite{BER-MIX-STO-2006}:

$$\textup{d}_\textsc{dcj}(\mathbb{S}_1, \mathbb{S}_2)=n_*-\left(c+\frac{p_\textup{e}}{2}\right)=n_*-\left(c_2 + c_4 + \ldots + c_{\infty} + \frac{p_0+p_2+\ldots+p_\infty}{2}\right).$$

If $\mathbb{S}_1=\{(\T{1}\z\rev{\T{3}}\z\T{2})\z[\T{4}]\}$ and $\mathbb{S}_2=\{(\T{1}\z\T{2})\z[\T{3}\z\rev{\T{4}}]\}$, then $n_*=4$, $c=1$ and $p_\textup{e}=2$ (see Figure~\ref{fig:bg}). Consequently, their DCJ distance is $\textup{d}_\textsc{dcj}(\mathbb{S}_1, \mathbb{S}_2)=2$.

\subsubsection{The class of $\sigma_k$ distances.}
Given the breakpoint graph of two canonical genomes $\mathbb{S}_1$ and $\mathbb{S}_2$, for $k \in \{2,4,6,\ldots,\infty\}$, we denote by $\sigma_k$ the cumulative sums $\sigma_k=c_2+c_4+\ldots+c_k+\frac{p_0+p_2+\ldots+p_{k-2}}{2}$. Then the $\sigma_k$ distance of $\mathbb{S}_1$ and $\mathbb{S}_2$ is defined to be~\cite{Chauve2018}:
$$\textup{d}_{\sigma_k}(\mathbb{S}_1, \mathbb{S}_2) = n_* - \sigma_k.$$

It is easy to see that the $\sigma_2$ distance equals the breakpoint distance and that the $\sigma_\infty$ distance equals the DCJ distance, and that the distance decreases monotonously between these two extremes.
Moreover, the $\sigma_k$ distance of two genomes that form a canonical pair can easily be computed in linear time for any $k \geq 2$.

\subsection{Comparing a singular and a duplicated genome}

Let $\mathbb{S}$ be a singular and $\mathbb{D}$ be a duplicated genome over the same $n_*$ gene families, that is, $\mathcal{F}(\mathbb{S})=\mathcal{F}(\mathbb{D})$ and $n_* = |\mathcal{F}(\mathbb{S})| =| \mathcal{F}(\mathbb{D})|$.
The number of genes in $\mathbb{D}$ is twice the number of genes in $\mathbb{S}$ and we need to somehow equalize the contents of these genomes, before
searching for common adjacencies and common telomeres of $\mathbb{S}$ and $\mathbb{D}$ or transforming one genome into the other with DCJ operations.
This can be done by \emph{doubling}~$\mathbb{S}$, with a rearrangement operation mimicking a \emph{whole genome duplication}: it simply consists of doubling each adjacency and each telomere of~$\mathbb{S}$.
However, when~$\mathbb{S}$ has one or more circular chromosomes, it is not possible to find a unique layout of its chromosomes after the doubling: indeed, each circular chromosome can be doubled into two identical circular chromosomes, or the two copies are concatenated to each other in a single circular chromosome.
Therefore, in general the doubling of a genome  $\mathbb{S}$ results in a set of doubled genomes denoted by $\mathtt{2}\mathbb{S}$.
Note that $|\mathtt{2}\mathbb{S}|=2^{r}$, where $r$ is the number of circular chromosomes in $\mathbb{S}$.
For example, if $\mathbb{S}=\{(\mathtt{1\,2})\,[\T{3}\,\T{4}]\}$, then $\mathtt{2}\mathbb{S}=\{\mathbb{B}_1,\mathbb{B}_2\}$ with
$\mathbb{B}_1=\{(\mathtt{1\,2})\,(\mathtt{1\,2})\,[\T{3}\,\T{4}]\,[\T{3}\,\T{4}]\}$ and
$\mathbb{B}_2=\{(\mathtt{1\,2\,1\,2})\,[\T{3}\,\T{4}]\,[\T{3}\,\T{4}]\}$ (see Table~\ref{tab:genome-types}).
All genomes in $\mathtt{2}\mathbb{S}$ have exactly the same multisets of adjacencies and of telomeres, therefore we can use a special notation for these multisets: $\mathcal{A}(\mathtt{2}\mathbb{S})=\mathcal{A}(\mathbb{S})\!\cup\!\mathcal{A}(\mathbb{S})$ and $\mathcal{T}(\mathtt{2}\mathbb{S})=\mathcal{T}(\mathbb{S})\!\cup\!\mathcal{T}(\mathbb{S})$.

Each family in a duplicated genome can be $\binom{\mathtt{a}}{\mathtt{b}}$-\emph{singularized} by adding the index $\mathtt{a}$ to one of its occurrences and the index $\mathtt{b}$ to the other. A duplicated genome can be entirely singularized if each of its families is singularized.
Let $\mathfrak{S}^\mathtt{a}_\mathtt{b}(\mathbb{D})$ be the set of all possible genomes obtained by all distinct ways of $\binom{\mathtt{a}}{\mathtt{b}}$-singularizing the duplicated genome~$\mathbb{D}$. Similarly, we denote by $\mathfrak{S}^\mathtt{a}_\mathtt{b}(\mathtt{2}\mathbb{S})$ the set of all possible genomes obtained by all distinct ways of $\binom{\mathtt{a}}{\mathtt{b}}$-singularizing each doubled genome in the set $\mathtt{2}\mathbb{S}$. 

\subsubsection{The class of $\sigma_k$ double distances.} 

The
class of $\sigma_k$ double distances of a singular genome $\mathbb{S}$ and duplicated genome $\mathbb{D}$ for $k=2,4,6,\ldots$ is defined as follows:
$$\textup{d}^2_{\sigma_k}(\mathbb{S}, \mathbb{D}) =\textup{d}_{\sigma_k}^2(\mathbb{S},\check{\mathbb{D}})=\min_{\mathbb{B}\in\mathfrak{S}^\mathtt{a}_\mathtt{b}(\mathtt{2}\mathbb{S})}\{ \textup{d}_{\sigma_k}(\mathbb{B},\check{\mathbb{D}})\}, \text{ where } \check{\mathbb{D}} \text{ is any genome in } \mathfrak{S}^\mathtt{a}_\mathtt{b}(\mathbb{D}).$$

\noindent
Observe that $\textup{d}_{\sigma_k}^2(\mathbb{S},\check{\mathbb{D}})=\textup{d}_{\sigma_k}^2(\mathbb{S},\check{\mathbb{D}}')$ 
for any $\check{\mathbb{D}}, \check{\mathbb{D}}' \in \mathfrak{S}^\mathtt{a}_\mathtt{b}(\mathbb{D})$.


\subsubsection{$\sigma_2$ (breakpoint) double distance.}

The \emph{breakpoint double distance} of $\mathbb{S}$ and~$\mathbb{D}$, denoted by $\textup{d}_\textsc{bp}^2(\mathbb{S},\mathbb{D})$, is equivalent to the $\sigma_2$ double distance.
For this case the solution can be found easily with a greedy algorithm~\cite{TAN-ZHE-SAN-2009}: each adjacency or telomere of $\mathbb{D}$ that occurs in $\mathbb{S}$ can be fulfilled. If an adjacency or telomere that occurs twice in $\mathbb{D}$ also occurs in~$\mathbb{S}$, it can be fulfilled twice in any genome from $\mathtt{2}\mathbb{S}$. 
Then,
$$\textup{d}_\textsc{bp}^2(\mathbb{S},\mathbb{D})=2n_*-|\mathcal{A}(\mathtt{2}\mathbb{S}) \cap \mathcal{A}(\mathbb{D})|-\frac{|\mathcal{T}(\mathtt{2}\mathbb{S}) \cap \mathcal{T}(\mathbb{D})|}{2}.$$

\subsubsection{$\sigma_\infty$ (DCJ) double distance.}
For the \emph{DCJ double distance}, that is equivalent to the $\sigma_\infty$ double distance, the solution space cannot be explored greedily. In fact, computing the DCJ double distance of
genomes $\mathbb{S}$ and $\mathbb{D}$ was proven to be an NP-hard problem~\cite{TAN-ZHE-SAN-2009}. 

\subsubsection{The complexity of $\sigma_k$ double distances.}
The exploration of the complexity space between the greedy linear time $\sigma_2$ (breakpoint) double distance and the NP-hard $\sigma_\infty$ (DCJ) double distance is the main motivation of this study. In the remainder of this paper we show that both $\sigma_4$ and $\sigma_6$ double distances can be solved in linear time. 

\section{Equivalence of \boldmath$\sigma_k$\,double\,distance and $\sigma_k$\,disambiguation}

A nice way of representing the solution space of the $\sigma_k$ double distance is by using a modified version of the breakpoint graph~\cite{TAN-ZHE-SAN-2009}.

\subsection{Ambiguous breakpoint graph}
Given a singular genome $\mathbb{S}$ and a duplicated genome $\mathbb{D}$, their \emph{ambiguous breakpoint graph} $ABG(\mathbb{S}, \check{\mathbb{D}}) = (V,E)$ is a multigraph representing the adjacencies of any element in $\mathfrak{S}^\mathtt{a}_\mathtt{b}(\mathtt{2}\mathbb{S})$ and a genome $\check{\mathbb{D}} \in \mathfrak{S}^\mathtt{a}_\mathtt{b}(\mathbb{D})$.
The vertex set $V$ comprises, for each family $\mathtt{X}$ in $\mathcal{F}(\mathbb{S})$, the two pairs of \emph{paralogous vertices} $\mathtt{X}_\mathtt{a}^{h}$, $\mathtt{X}_\mathtt{b}^{h}$ and $\mathtt{X}_\mathtt{a}^{t}$, $\mathtt{X}_\mathtt{b}^{t}$.
We can use the notation~$\hat{u}$ to refer to the paralogous counterpart of a vertex $u$. For example, if $u=\mathtt{X}_\mathtt{a}^{h}$, then $\hat{u}=\mathtt{X}_\mathtt{b}^{h}$.

\begin{figure}[ht!]
  
  \begin{center}
  \includegraphics[scale=0.36]{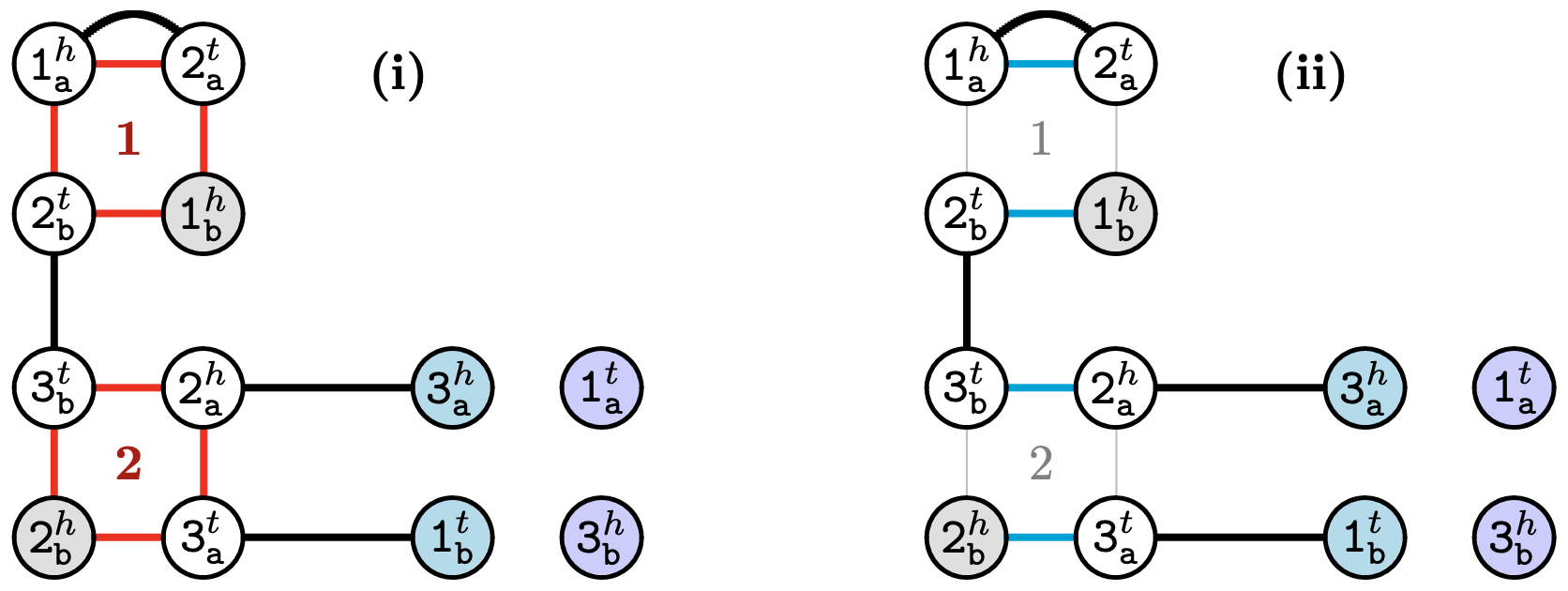}
  \end{center}
  \caption{\label{fig:abg}(i) Ambiguous breakpoint graph $ABG(\mathbb{S}, \check{\mathbb{D}})$ for genomes $\mathbb{S}=\{[\mathtt{1\,2\,3}]\}$ and $\check{\mathbb{D}}=\{[\mathtt{1_a\,2_a}\,\overline{\mathtt{3}}_\mathtt{a}\,\mathtt{1_b}]\,[\overline{\mathtt{3}}_\mathtt{b}\,\mathtt{2_b}]\}$. Edge types are distinguished by colors: $\check{\mathbb{D}}$-edges are drawn in black and $\mathbb{S}$-edges (squares) are drawn in red. (ii) Induced breakpoint graph $BG(\tau,\check{\mathbb{D}})$ in which all squares are resolved by the solution $\tau=(\{\mathtt{1}_\mathtt{a}^h\mathtt{2}_\mathtt{a}^t,\mathtt{1}_\mathtt{b}^h\mathtt{2}_\mathtt{b}^t\},\{\mathtt{2}_\mathtt{a}^h\mathtt{3}_\mathtt{b}^t,\mathtt{2}_\mathtt{b}^h\mathtt{3}_\mathtt{a}^t\}\})$, resulting in one 2-cycle, two 0-paths, one 2-path and one 4-path. This is also the breakpoint graph of $\check{\mathbb{D}}$ and $\mathbb{B}=\{[\mathtt{1_a\,2_a\,3_b}],[\mathtt{1_b\,2_b\,3_a}]\} \in \mathfrak{S}^\mathtt{a}_\mathtt{b}(\mathtt{2}\mathbb{S})$. In both (i) and (ii), vertex types are distinguished by colors: telomeres in $\g{S}$ are marked in blue, telomeres in $\check{\g{D}}$ are marked in gray, telomeres in both $\g{S}$ and $\check{\g{D}}$ are marked in purple and non-telomeric vertices are white.}
  
\end{figure}

The edge set $E$ represents the adjacencies.
For each adjacency in $\check{\mathbb{D}}$ there exists one $\check{\mathbb{D}}$-edge in $E$ linking its two extremities.
The $\mathbb{S}$-edges represent all adjacencies occurring in all genomes from $\mathfrak{S}^\mathtt{a}_\mathtt{b}(\mathtt{2}\mathbb{S})$:
for each adjacency $\upgamma\upbeta$ of $\mathbb{S}$, we have the \emph{pair of paralogous edges} $\mathcal{E}(\upgamma\upbeta)=\{\upgamma_\mathtt{a}\upbeta_\mathtt{a}, \upgamma_\mathtt{b}\upbeta_\mathtt{b}\}$ and the \emph{complementary pair of paralogous edges} $\Tilde{\mathcal{E}}(\upgamma\upbeta)=\{\upgamma_\mathtt{a}\upbeta_\mathtt{b}, \upgamma_\mathtt{b}\upbeta_\mathtt{a}\}$.
Note that $\TTilde{\mathcal{E}}(\upgamma\upbeta)=\mathcal{E}(\upgamma\upbeta)$. 
The \emph{square} of $\upgamma\upbeta$ is then $\mathcal{Q}(\upgamma\upbeta)=\mathcal{E}(\upgamma\upbeta)\cup\Tilde{\mathcal{E}}(\upgamma\upbeta)$.
The $\mathbb{S}$-edges in the ambiguous breakpoint graph are therefore the squares of all adjacencies in $\mathbb{S}$. 
Let $a_*$ be the number of squares in $ABG(\mathbb{S}, \check{\mathbb{D}})$.
Obviously we have $a_*=|\mathcal{A}(\g{S})|=n_*-\kappa(\g{S})$, where $\kappa(\g{S})$ is the number of linear chromosomes in $\g{S}$.
Again, we can use the notation $\hat{e}$ to refer to the paralogous counterpart of an $\mathbb{S}$-edge $e$.
For example, if $e=\upgamma_\mathtt{a}\upbeta_\mathtt{a}$, then $\hat{e}=\upgamma_\mathtt{b}\upbeta_\mathtt{b}$.
An example of an ambiguous breakpoint graph is shown in Figure~\ref{fig:abg}~(i).

Each linear chromosome in $\g{S}$ corresponds to four telomeres, called \emph{$\g{S}$-telomeres}, in any element of $\T{2}\g{S}$. These four vertices are not part of any square. In other words, the number of $\g{S}$-telomeres in $ABG(\mathbb{S}, \check{\mathbb{D}})$ is $4\kappa(\g{S})$. If $\kappa(\g{D})$ is the number of linear chromosomes in $\g{D}$, the number of telomeres in $\check{\g{D}}$, also called \emph{$\check{\g{D}}$-telomeres}, is~$2\kappa(\g{D})$. 

\subsection{The class of $\sigma_k$ disambiguations}

\emph{Resolving} a square $\mathcal{Q}(\cdot)=\mathcal{E}(\cdot)\cup\Tilde{\mathcal{E}}(\cdot)$ corresponds to \emph{choosing} in the ambiguous breakpoint graph either the edges from~$\mathcal{E}(\cdot)$ or the edges from $\Tilde{\mathcal{E}}(\cdot)$, while the complementary pair is \emph{masked}. Resolving all squares is called \emph{disambiguating} the ambiguous breakpoint graph. If we number the squares of $ABG(\mathbb{S},\check{\mathbb{D}})$ from~1 to~$a_*$, a \emph{solution} can be represented by a tuple
$\tau=(\mathcal{L}_1,\mathcal{L}_2, \ldots, \mathcal{L}_{a_*})$, where each $\mathcal{L}_i$ contains the pair of paralogous edges (either~$\mathcal{E}_i$ or~$\Tilde{\mathcal{E}}_i$) that are chosen (kept) in the graph for square $\mathcal{Q}_i$. The graph \emph{induced} by $\tau$ is a simple breakpoint graph,
which we denote by $BG(\tau,\check{\mathbb{D}})$.
Figure~\ref{fig:abg}~(ii) shows an example.

Given a solution $\tau$, let $c_i$ and $p_j$ be, respectively, the number of cycles of length $i$ and of paths of length $j$ in $BG(\tau,\check{\mathbb{D}})$. The $k$-score of $\tau$ is then the sum $\sigma_k=c_2+c_4+\ldots+c_k+\frac{p_0+p_2+\ldots+p_{k-2}}{2}$.
The minimization problem of computing the $\sigma_k$ double distance of $\mathbb{S}$ and $\mathbb{D}$ is equivalent to finding a solution~$\tau$ so that the $k$-score of $\tau$ is maximized~\cite{TAN-ZHE-SAN-2009}. We call the latter (maximization) problem \emph{$\sigma_k$ disambiguation}.
As already mentioned, for $\sigma_2$ the double distance can be solved in linear time  and for $\sigma_\infty$ the double distance is NP-hard. Therefore the same is true, respectively, for the $\sigma_2$ and the $\sigma_\infty$ disambiguations. Conversely, if we determine the complexity of solving the $\sigma_k$ disambiguation for any $k \geq 4$, this will automatically determine the complexity of solving the $\sigma_k$ double distance.

An optimal solution for the $\sigma_k$ disambiguation of $ABG(\mathbb{S},\check{\mathbb{D}})$ gives its \emph{$k$-score}, denoted by $\sigma_k(ABG(\mathbb{S},\check{\mathbb{D}}))$.
Note that, since an optimal $\sigma_k$ disambiguation is also a $\sigma_{k+2}$ disambiguation, although possibly not optimal, the $k$-score of $ABG(\mathbb{S},\check{\mathbb{D}})$ can not decrease as $k$ increases.

\subsubsection{Approach for solving the $\sigma_k$ disambiguation.}

A \emph{player} of the $\sigma_k$ disambiguation is either a
valid cycle whose length is at most~$k$ or a valid even path whose length is at most $k-2$.
In order to solve the $\sigma_k$ disambiguation, 
a natural approach is to visit $ABG(\mathbb{S},\check{\mathbb{D}})$ and search for players.
For describing how the graph can be screened, we need to introduce the following concepts.
Two $\mathbb{S}$-edges in $ABG(\mathbb{S},\check{\mathbb{D}})$ are \emph{incompatible} when they belong to the same square and are not paralogous.
A component in $ABG(\mathbb{S},\check{\mathbb{D}})$ is \emph{valid} when it does not contain any pair of incompatible edges.
Note that a valid component necessarily alternates $\mathbb{S}$-edges and $\check{\mathbb{D}}$-edges.
Two valid components $C\ne C'$ in $ABG(\mathbb{S},\check{\mathbb{D}})$ are either \emph{intersecting}, when they share at least one vertex, or \emph{disjoint}. 
It is obvious that any solution $\tau$ of $ABG(\mathbb{S},\check{\mathbb{D}})$ is composed of disjoint valid components.

Given a solution $\tau=(\mathcal{L}_1,\mathcal{L}_2, \ldots, \mathcal{L}_i \ldots, \mathcal{L}_{a_*})$, the \emph{switching} operation of the $i$-th element of $\tau$ is denoted by $\Tilde{s}(\tau,i)$ and replaces value $\mathcal{L}_i$ by $\Tilde{\mathcal{L}_i}$ resulting in $\tau'=(\mathcal{L}_1,\mathcal{L}_2, \ldots, \Tilde{\mathcal{L}_i} \ldots, \mathcal{L}_{a_*})$.
A choice of paralogous edges resolving a given square $\mathcal{Q}_i$ can be \emph{fixed} for any solution, meaning that 
$\mathcal{Q}_i$ can no longer be switched. In this case, $\mathcal{Q}_i$ is itself said to be \emph{fixed}.

\section{First steps to solve the \boldmath$\sigma_k$ disambiguation}

In this section we describe a greedy linear time algorithm for the $\sigma_4$ disambiguation and give some general results related to any $\sigma_k$ disambiguation.

\subsection{Common adjacencies and telomeres are conserved}

Let $\tau$ be an optimal solution for $\sigma_k$ disambiguation of $ABG(\mathbb{S},\check{\mathbb{D}})$. 
If a player $C\in BG(\tau,\check{\mathbb{D}})$ is disjoint from any player distinct from $C$ in any other optimal solution,
then $C$ must be part of all optimal solutions and is itself said to be \emph{optimal}.

\begin{lemma}\label{lemma:non-conflicting-2-cycles}
For any $\sigma_k$ disambiguation, all existing 0-paths and 2-cycles in $ABG(\mathbb{S},\mathbb{D})$ are optimal.
\end{lemma}

\begin{figure}[ht!]
    \begin{center}
  \includegraphics[scale=0.36]{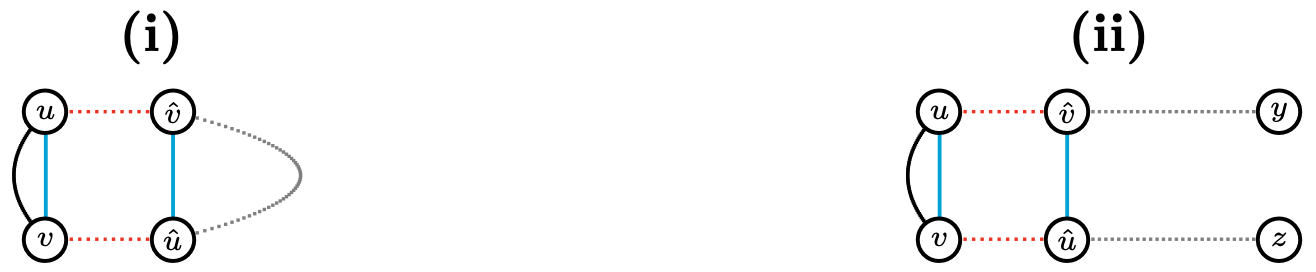}
  \end{center}
    
    \caption{(i)
    The gray path connecting vertices $\hat{v}$ to~$\hat{u}$ is necessarily odd with length at least one and alternates 
    $\check{\mathbb{D}}$- and 
    $\mathbb{S}$-edges. The 2-cycle $C=(uv)$ intersects the longer cycle $D=(u\hat{v}\ldots \hat{u}v)$. Any solution containing (red edges) $\Tilde{\mathcal{E}}=\{u\hat{v}, \hat{u}v\}$ induces~$D$ and can be improved by switching $\Tilde{\mathcal{E}}$ to 
    (blue edges)
    $\mathcal{E}=\{uv, \hat{u}\hat{v}\}$, inducing, instead of~$D$, the 2-cycle $C$ and cycle $D'=(\hat{v}\ldots \hat{u})$ (which is shorter than $D$). (ii)
    The gray paths connecting vertices $\hat{v}$ to telomere~$y$ and~$\hat{u}$ to telomere~$z$ alternate 
    $\check{\mathbb{D}}$- and 
    $\mathbb{S}$-edges. The 2-cycle $C=(uv)$ intersects the longer path $P=y\ldots\hat{v}uv\hat{u}\ldots z$. Any solution containing (red edges) $\Tilde{\mathcal{E}}=\{u\hat{v}, \hat{u}v\}$ induces~$P$ and can be improved by switching $\Tilde{\mathcal{E}}$ to 
    (blue edges)
    $\mathcal{E}=\{uv, \hat{u}\hat{v}\}$, inducing, instead of~$P$, the 2-cycle $C$ and path $P'=y\ldots\hat{v}\hat{u}\ldots z$ (which is of the same type, but 2 edges shorter than $P$).}
    \label{fig:2-cycle}
\end{figure}

\proof While any 0-path is an isolated vertex and obviously optimal, the optimality of every 2-cycle is less obvious but still holds, as illustrated in Figure~\ref{fig:2-cycle}. \qed

This lemma is a generalization of the (breakpoint) $\sigma_2$ disambiguation and guarantees that all common adjacencies and telomeres are conserved in any $\sigma_k$ double distance, including the NP-hard (DCJ) $\sigma_\infty$ case.
All 0-paths are isolated vertices that do not integrate squares, therefore they are selected independently of the choices for resolving the squares. A 2-cycle, in its turn, always includes one $\mathbb{S}$-edge from some square (such as square~1 in Figure~\ref{fig:abg}).
From now on we assume that 
squares that have at least one $\mathbb{S}$-edge in a 2-cycle
are fixed so that all existing 2-cycles are induced.

\subsection{Symmetric squares can be fixed arbitrarily}

\begin{figure}[ht!]
    \begin{center}
  \includegraphics[scale=0.36]{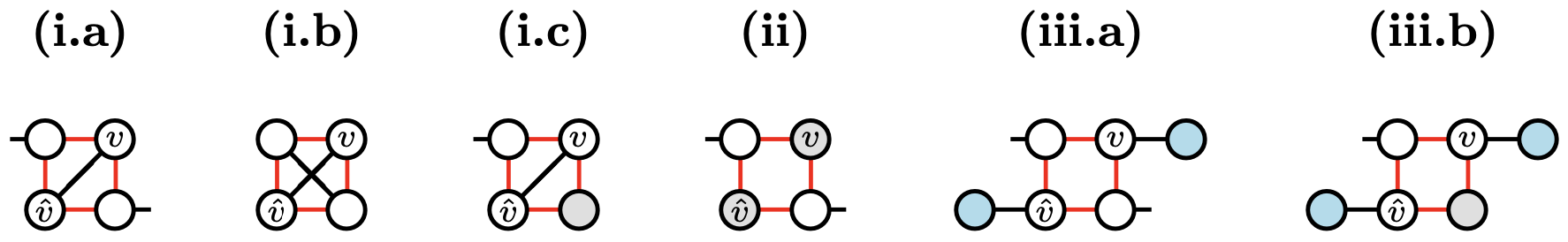}
\end{center}
\caption{Possible symmetric squares in the ambiguous breakpoint graph.}\label{fig:symmetric-squares}

\end{figure}

Let a \emph{symmetric square} in $ABG(\mathbb{S},\check{\mathbb{D}})$ either (i) have a $\check{\g{D}}$-edge connecting a pair of paralogous vertices,
or (ii) have $\check{\g{D}}$-telomeres in one pair of paralogous vertices, or (iii) have $\check{\g{D}}$-edges directly connected to $\g{S}$-telomeres inciding in one pair of paralogous vertices, as illustrated in Figure~\ref{fig:symmetric-squares}. Note that, for any $\sigma_k$ disambiguation, the two ways of resolving each of these squares would lead to solutions with the same score, therefore each of them can be fixed arbitrarily. From now on we assume that $ABG(\mathbb{S},\check{\mathbb{D}})$ has no symmetric squares.

\subsection{A linear time greedy algorithm for the $\sigma_4$ disambiguation}

\begin{figure}[ht!]
 \begin{center}
  \includegraphics[scale=0.36]{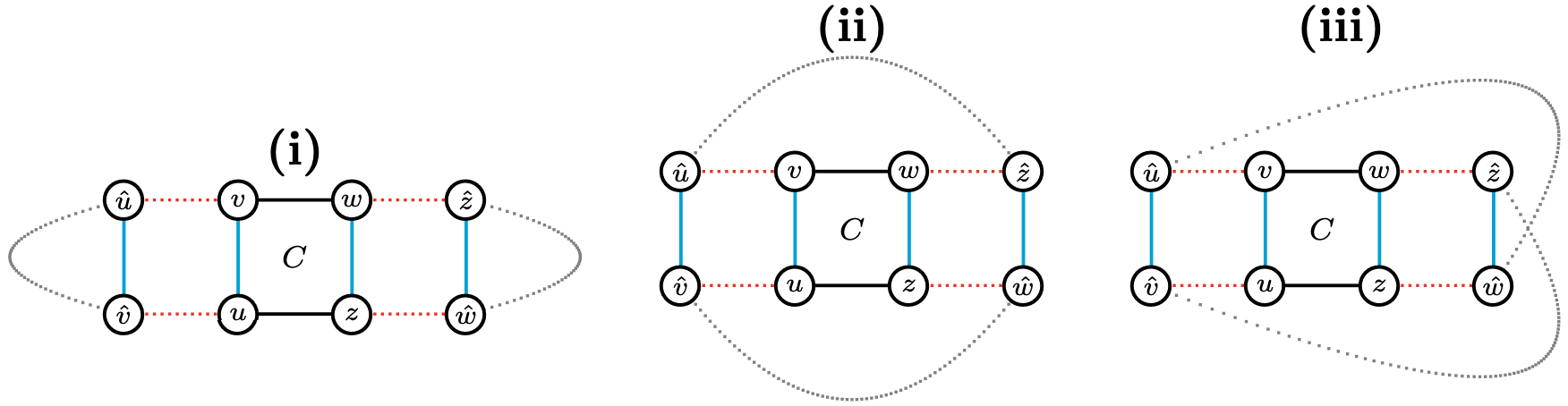}
\end{center}
    
    \caption{\label{fig:4-cycle}Illustration of the co-optimality of 
    every valid 4-cycle not intersecting a 2-cycle in the $\sigma_4$ disambiguation. 
    In each of these pictures, each gray path is necessarily odd with length at least one and alternates 
    $\check{\mathbb{D}}$- and 
    $\mathbb{S}$-edges. Furthermore, the 4-cycle $C=(uvwz)$ is displayed in the center, induced by blue edges.
    In (i) it is easy to see that any optimal solution is induced by the blue edges and includes, besides the cycle~$C$, cycles $(\hat{u}\ldots\hat{v})$ and $(\hat{w}\ldots\hat{z})$.
    In (ii) an optimal solution includes 4-cycle $C$ and cycle $C'=(\hat{u}\hat{v}\ldots\hat{w}\hat{z}\ldots)$. If the connection between $\hat{v}$ and $\hat{w}$ is a single edge, then another optimal solution is induced by the red edges, including 4-cycle $D=(u\hat{v}\hat{w}z)$ and cycle $D'=(v\hat{u}\ldots\hat{z}w)$. And if additionally the connection between $\hat{u}$ and $\hat{z}$ is a single edge, then both~$C'$ and~$D'$ are also 4-cycles. In (iii) any optimal solution is induced by the blue edges and includes 4-cycle $C$ and cycle $(\hat{u}\hat{v}\ldots\hat{z}\hat{w}\ldots)$, which is also a 4-cycle when the connections between $\hat{v}$ and $\hat{z}$ and between $\hat{u}$ and $\hat{w}$ are single edges.} 
\end{figure}

Differently from 2-cycles, two valid 4-cycles can intersect with each other. But, since our graph is free of symmetric squares, two valid 2-paths cannot intersect with each other.
Moreover, since a 2-path has no $\g{\check{D}}$-edge connecting squares, a 4-cycle and a 2-path cannot intersect with each other.
In this setting, it is clear that, for the $\sigma_4$ disambiguation, any valid 2-path is always optimal. Furthermore, a 4-cycle that does intersect with another one is always optimal and two intersecting 4-cycles are always part of two co-optimal solutions:

\begin{lemma}\label{lemma:4-cycles-in-c4}
Any valid 4-cycle that is disjoint from a 2-cycle in $ABG(\mathbb{S},\mathbb{D})$ is induced by an optimal solution of $\sigma_4$ disambiguation.
\end{lemma}
\proof
All possible patterns are represented in Figure~\ref{fig:4-cycle}: A valid 4-cycle $C$ (in the center) connecting two squares and the three distinct possibilities of linking the four open ends. In all cases the valid 4-cycle~$C$ is either optimal or co-optimal.
\qed

An optimal solution of $\sigma_4$ disambiguation can then be  obtained greedily:
after fixing squares containing edges that are part of 2-cycles, traverse the remainder of the graph and, for each valid 2-path or 4-cycle $C$ that is found, 
fix the square(s) containing $\mathbb{S}$-edges that are part of $C$, so that $C$ is induced. When this part is accomplished the remaining squares can be fixed arbitrarily.

\subsection{Pruning $ABG(\mathbb{S},\check{\mathbb{D}})$ for the $\sigma_6$ disambiguation} 

A player in the $\sigma_6$ disambiguation can be either a \emph{$\ppl$-path}, that is a valid 2- or 4-path, or a \emph{$\cpl$-cycle}, that is  a valid 4- or 6-cycle.  It is easy to see that players can intersect with each other. 
Moreover, for the $\sigma_6$ disambiguation, not every player is induced by at least one optimal solution. For that reason, a greedy algorithm does not work here and a more elaborated procedure is required.
The first step is a linear time preprocessing in which from $ABG(\mathbb{S},\check{\mathbb{D}})$ first all edges are removed that are incompatible with the existing 2-cycles, and then all remaining edges that cannot be part of a player.
This results in a \emph{$\{6\}$-pruned} ambiguous breakpoint graph $PG(\mathbb{S},\check{\mathbb{D}})$. 

The first step is easily achieved by a simple graph traversal in which for each $\check{\mathbb{D}}$-edge $uv$ it is tested whether both ends connect to the same $\mathbb{S}$-edge $uv$.
If this is the case, the two incident $\mathbb{S}$-edges $u\hat v$ and $v\hat u$ are removed from the graph, separating the 2-cycle $(uv)$.
Then, in the second step, for any remaining edge $e$, its 6-neighborhood 
(which has constant size in a graph of degree at most three) is exhaustively explored for the existence of a player involving~$e$.
If no such player is found, $e$ is deleted.
Each of these two steps clearly takes linear time $O(|ABG(\mathbb{S},\check{\mathbb{D}})|)$, and what remains is exactly the desired graph $PG(\mathbb{S},\check{\mathbb{D}})$.

\begin{figure}[ht!]
 \begin{center}
  \includegraphics[scale=0.36]{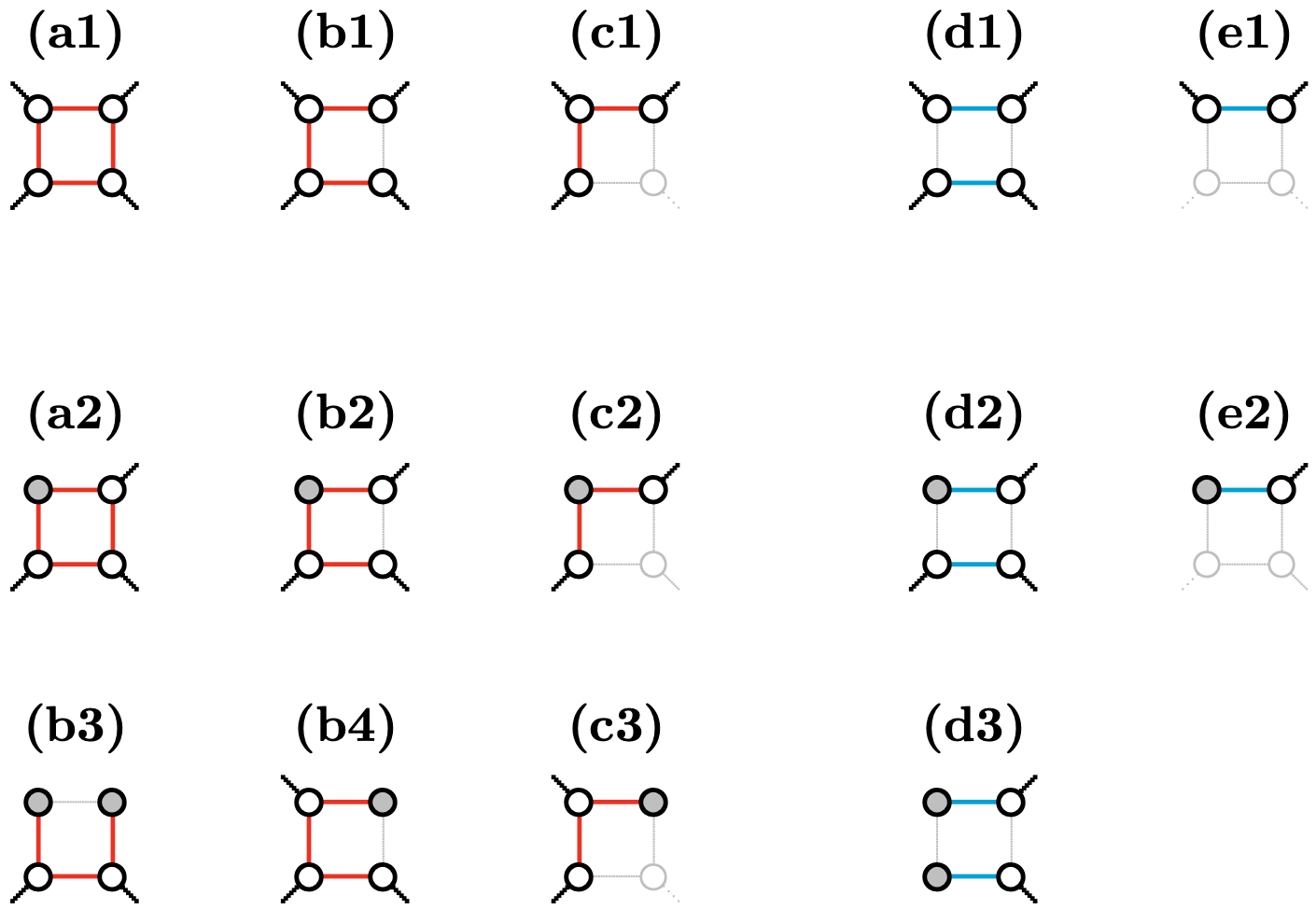}
  \end{center}
   
     \caption{Possible (partial) squares of $PG(\mathbb{S},\check{\mathbb{D}})$. Shadowed parts represent the pruned elements
     (since they do not count for the score, it is not relevant to differentiate whether the pruned vertices are telomeres or not). 
     The top line represents squares whose preserved elements include no telomere. The middle and the bottom line represent squares whose preserved elements include telomeres, marked in gray. Note that all of these are  $\check{\mathbb{D}}$-telomeres ($\mathbb{S}$-telomeres are not part of any square).
     Cases (a1-a2), (b1-b4) and (c1-c3) are ambiguous, while cases (d1-d3) and (e1-e2) are resolved.
     }
    \label{fig:partial-sq}
\end{figure}

The edges that are not pruned and are therefore present in $PG(\mathbb{S},\check{\mathbb{D}})$ are said to be \emph{preserved}.
As shown in Figure~\ref{fig:partial-sq}, for any given square the pruned graph might preserve either (a1-a2) all edges, or (b1-b4) only three edges, or (c1-c3) only two edges each one from a distinct pair of paralogous edges, or (d1-d3) only two edges from the same pair of paralogous edges, or (e1-e2) a single edge. 
While the squares are still ambiguous in cases (a1-a2), (b1-b4) and (c1-c3), in cases (d1-d3) and (e1-e2) they are already resolved and can be fixed according to the preserved paralogous edges in cases (d1-d3) and (e1-e2).
Additionally, if none of its edges is part of a player, a square is completely pruned out and is arbitrarily fixed in $ABG(\mathbb{S},\check{\mathbb{D}})$.

The smaller pruned graph $PG(\mathbb{S},\check{\mathbb{D}})$ has all relevant parts required for finding an optimal solution of $\sigma_6$ disambiguation, therefore the $6$-scores of both graphs are the same:  $\sigma_6(ABG(\mathbb{S},\check{\mathbb{D}}))=\sigma_6(PG(\mathbb{S},\check{\mathbb{D}}))$.
A clear advantage here is that the pruned graph might be split into smaller connected components, and it is obvious that the disambiguation problem can be  solved independently for each one of them. Any square that 
is still ambiguous in $PG(\mathbb{S},\check{\mathbb{D}})$ is called
a \emph{$\{6\}$-square}.
Each connected component $G$ of $PG(\mathbb{S},\check{\mathbb{D}})$ is of one of the two types:
\begin{enumerate}
\item \emph{Ambiguous}: $G$ includes at least one $\{6\}$-square; 
\item \emph{Resolved} \emph{(trivial)}: $G$ is either a simple valid 0-, 2- or 4-path or a simple valid 2-, 4- or 6-cycle.
\end{enumerate}

Let $\mathcal{C}$ and $\mathcal{P}$ be the sets of resolved components, so that $\mathcal{C}$ has all resolved cycles and $\mathcal{P}$ has all resolved paths. Furthermore, let $\mathcal{M}$ be the set of ambiguous components of $PG(\mathbb{S},\check{\mathbb{D}})$.
If we denote by $\sigma_6(M)$ the 6-score of an ambiguous component $M \in \mathcal{M}$, the 6-score of $PG(\mathbb{S},\check{\mathbb{D}})$ can be computed with the formula:
$$\sigma_6(PG(\mathbb{S},\check{\mathbb{D}}))=|\mathcal{C}|+\frac{|\mathcal{P}|}{2}+\sum_{M \in\z \mathcal{M}} \sigma_6(M).$$ 

Solving the $\sigma_6$ disambiguation corresponds then to finding, for each ambiguous component $M\!\in\!\mathcal{M}$, an optimal solution 
including only the $\{6\}$-squares of $M$.
From now on, by $\mathbb{S}$-edge, $\mathbb{S}$-telomere, $\check{\mathbb{D}}$-edge and $\check{\mathbb{D}}$-telomere, we are referring only to the elements that are preserved in $PG(\mathbb{S},\check{\mathbb{D}})$.

\section{Intersection between players of the \boldmath$\sigma_6$ disambiguation}

Let a \emph{$\check{\g{D}}\g{S}\check{\g{D}}$-path} be a subpath of three edges, starting and ending with a $\check{\g{D}}$-edge. This is the largest segment that can be shared by two players: although there is no room to allow distinct $\{2,4\}$-paths and/or valid 4-cycles to share a $\check{\g{D}}\g{S}\check{\g{D}}$-path in a graph free of symmetric squares, a $\check{\g{D}}\g{S}\check{\g{D}}$-path can be shared by at most two valid 6-cycles. Furthermore, if distinct $\check{\g{D}}\g{S}\check{\g{D}}$-paths intersect at the same $\check{\g{D}}$-edge $e$ and each of them occurs in two distinct 6-cycles, then the $\check{\g{D}}$-edge $e$ occurs in four distinct valid 6-cycles.

\begin{figure}[ht!]

\begin{center}
  \includegraphics[scale=0.36]{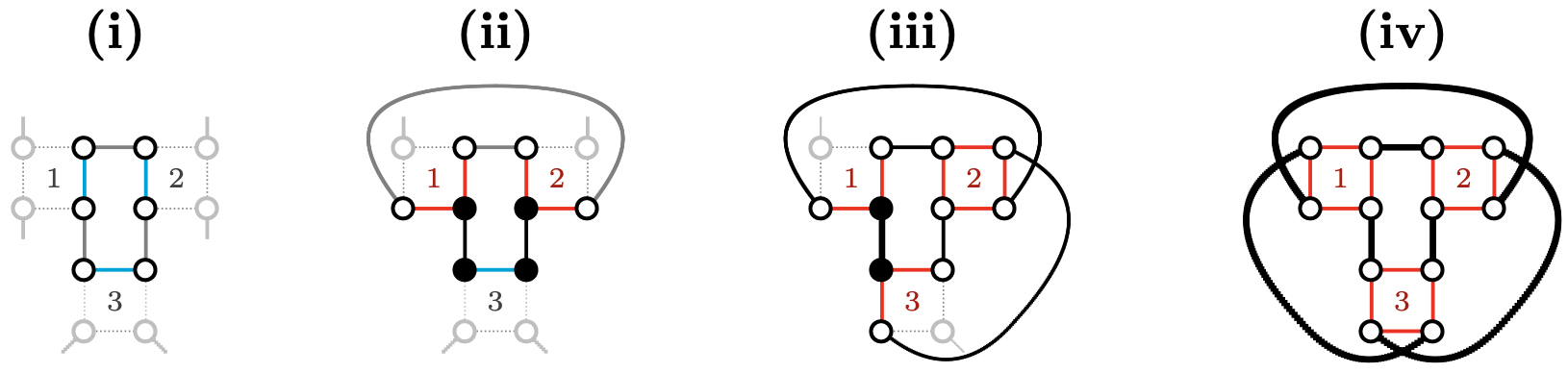}
  \end{center}
   
    \caption{(i) Resolved component ($\text{score}=1$): a 6-cycle alternating (black) $\check{\g{D}}$- and (blue) $\g{S}$-edges, without intersections. (ii) Two  6-cycles share one $\check{\g{D}}\g{S}\check{\g{D}}$-path composed of the two black $\check{\g{D}}$-edges with the blue $\g{S}$-edge in between. (iii) Unsaturated triplet with $\text{score}=1$: every $\check{\g{D}}\g{S}\check{\g{D}}$-path including the same $\check{\g{D}}$-edge (the thick black one) occurs in two distinct 6-cycles. The thick black $\check{\g{D}}$-edge occurs in four 6-cycles, all other black edges occur in two  6-cycles. (iv) Saturated triplet with $\text{score}=2$: every $\check{\g{D}}\g{S}\check{\g{D}}$-path  occurs in two distinct 6-cycles, every black edge occurs in four  6-cycles.}
    \label{fig:triplet}

\end{figure}

In Figure~\ref{fig:triplet} we characterize this exceptional situation, which consists of the occurrence of a \emph{triplet}, defined to be an ambiguous component composed of exactly three connected ambiguous squares in which at most two vertices, necessarily in distinct squares, are pruned out. In a \emph{saturated} triplet, the squares in each pair are connected to each other by two $\check{\g{D}}$-edges connecting paralogous vertices in both squares; if a single $\check{\g{D}}$-edge is missing, that is, the corresponding vertices have outer connections, we have an \emph{unsaturated} triplet. This structure and its score can be easily identified, therefore we will assume that our graph is free from triplets.
 With this condition, $\check{\g{D}}$-edges can be shared by at most two players:

\begin{proposition}\label{prop:Dedges}
Any $\check{\mathbb{D}}$-edge 
is part of either one or two (intersecting) players in a graph free of symmetric squares and triplets.
\end{proposition}

\proof
Recall that a \emph{$\check{\g{D}}\g{S}\check{\g{D}}$-path} is a subpath of three edges, starting and ending with a $\check{\g{D}}$-edge. It is easy to see that, without symmetric squares, there is no ``room'' to allow distinct 4-paths and/or 4-cycles to share a $\check{\g{D}}\g{S}\check{\g{D}}$-path. In contrast, at most two valid 6-cycles can share a $\check{\g{D}}\g{S}\check{\g{D}}$-path as illustrated in Figure~\ref{fig:triplet}. And if the $\g{S}$-edge in the middle of the shared $\check{\g{D}}\g{S}\check{\g{D}}$-path is in an ambiguous square, we have the exceptional case of a triplet, where a $\check{\g{D}}$-edge occurs in more than two players. This case can be treated separately in a preprocessing step, so that we can assume that our graph is free of triplets.

Let an \emph{$\g{S}\check{\g{D}}\g{S}$-path} be a subpath of three edges, starting and ending with an $\g{S}$-edge.
Obviously there is no ``room'' to allow two players to share an $\g{S}\check{\g{D}}\g{S}$-path: (i) there are two ways of adding a $\check{\g{D}}$-edge to a $\g{S}\check{\g{D}}\g{S}$-path for obtaining a valid 4-path but they are incompatible therefore at most one can exist; or (ii) the two ends of the $\g{S}\check{\g{D}}\g{S}$-path must incide in the same $\check{\g{D}}$-edge, giving a single way of obtaining a 4-cycle; 
 or (iii) any valid 6-cycle including the given $\g{S}\check{\g{D}}\g{S}$-path needs to have both extra $\check{\g{D}}$-edges inciding at both ends, then there can be only one way of filling the ``gap'' with a last $\g{S}$-edge.

Now let an \emph{open} 2-path be an $\g{S}$-edge adjacent to a $\check{\g{D}}$-edge such that at most one of the two includes a telomere. Considering the case of paths, 
in the absence of symmetric squares there is no possibility of having two 4-paths sharing an open 2-path. And considering the case of cycles, it is obvious that two $\{4,6\}$-cycles sharing the same open 2-path must share the same $\check{\g{D}}\g{S}\check{\g{D}}$-path, which falls in the same particular case of a triplet mentioned before.

\begin{figure}[ht!]
\begin{center}
  \includegraphics[scale=0.36]{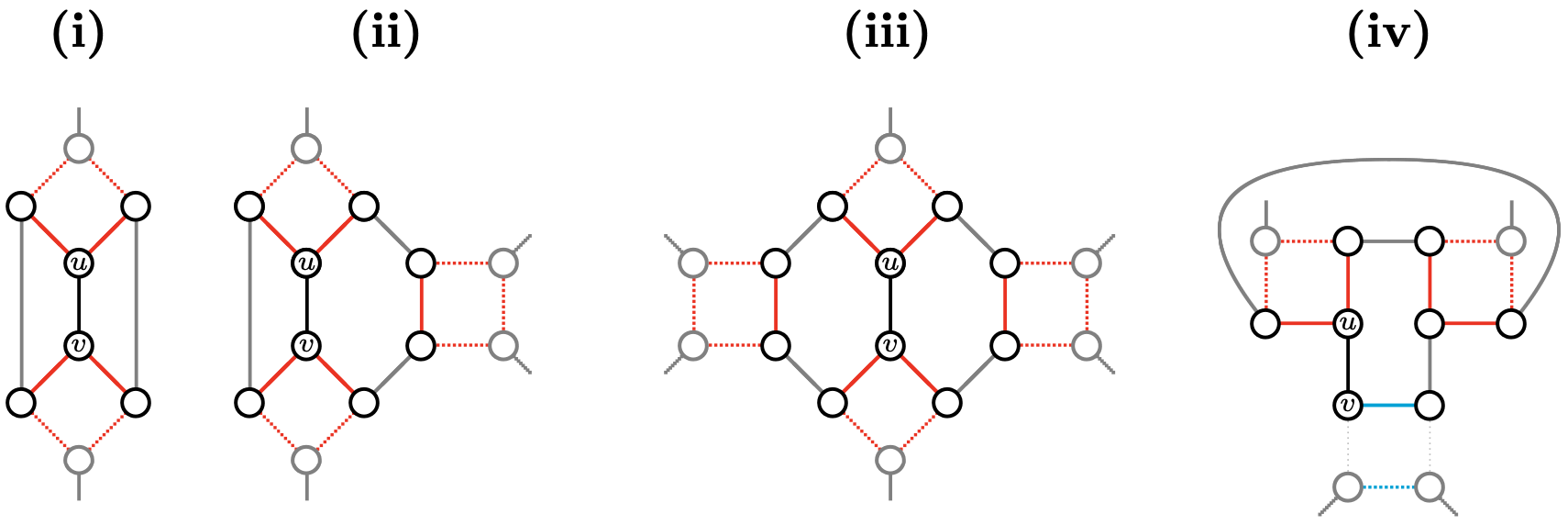}
  \end{center}
    
    \caption{Patterns free of triplets and symmetric squares showing a $\check{\mathbb{D}}$-edge $uv$ in two distinct intersecting $\cpl$-cycles which themselves do not intersect 2-cycles.  (i) -- (iii) The edge $uv$ connects two distinct squares and is part of two $\cpl$-cycles whose intersection is only~$uv$. (iv) The edge $uv$ is part of two 6-cycles whose intersection is a $\check{\mathbb{D}}\g{S}\check{\mathbb{D}}$-path starting in $uv$.  
    Here one square (marked in blue) is clearly fixed: if this square could be switched, this would merge each of the two existing 6-cycles into a longer cycle.}
    \label{fig:6-patterns}
\end{figure}

Finally, it is easy to see that a $\check{\g{D}}$-edge can occur in more than one player (general cases for cycles are illustrated in Figure~\ref{fig:6-patterns}). However, it can only occur in more than two players if it is part of distinct $\check{\g{D}}\g{S}\check{\g{D}}$-paths such that each of them occurs in distinct players. By construction we can see that this can only happen in a triplet (Figure~\ref{fig:triplet}) or if the graph has symmetric squares.
It follows that, without symmetric squares and triplets, each $\check{\g{D}}$-edge occurs in at most two distinct players.
\qed

\begin{proposition}\label{prop:Sedges}
	Any $\mathbb{S}$-edge of a $\{6\}$-square is part of exactly one player in a graph free of symmetric squares and triplets.
\end{proposition}
\proof
If an $\mathbb{S}$-edge $e$ is in a $\{6\}$-square $\mathcal{Q}$, it ``shares'' either the same $\check{\mathbb{D}}$-edge or the same $\check{\mathbb{D}}$-telomere~$d$ with another $\mathbb{S}$-edge $e'$ from the same square $\mathcal{Q}$. In this case the $\check{\mathbb{D}}$-edge/telomere $d$ is part of exactly two players and each of the $\mathbb{S}$-edges $e$ and $e'$ must be part of exactly one player.
\qed

In the next sections we present the most relevant contribution of this work: an algorithm to solve the~$\sigma_6$~disambiguation in linear time. 

\section{Solving the \boldmath$\sigma_6$ disambiguation for circular genomes}

For the case of \emph{circular genomes}, which are those exclusively including circular chromosomes, the ambiguous breakpoint graph has no telomeres, therefore all players are cycles. In this case, we call each ambiguous component a \emph{cycle-bubble}.

\begin{figure}[ht!]
    \begin{center}
  \includegraphics[scale=0.36]{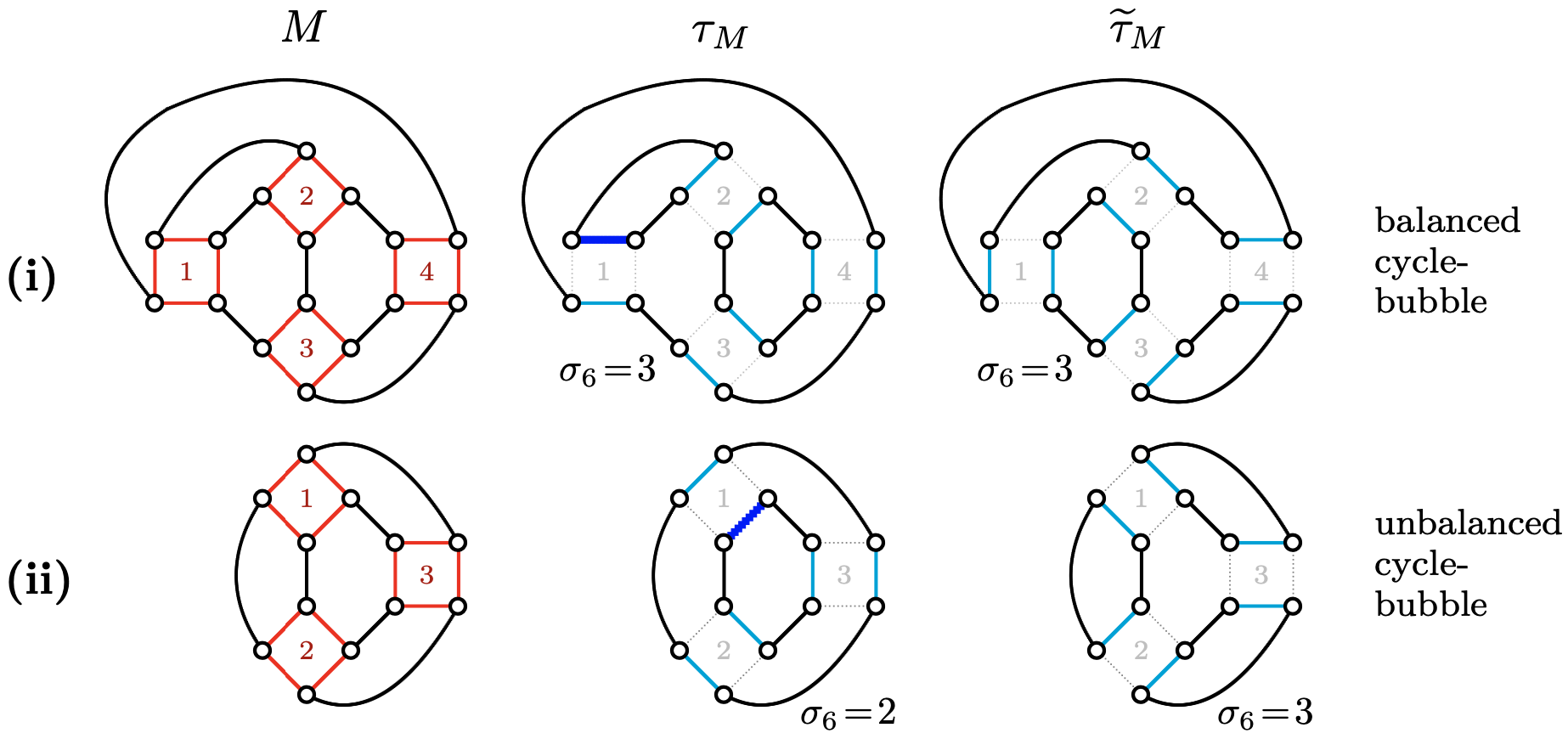}
  \end{center}
  
    \caption{Example of execution of Algorithm~\ref{alg:6-amb-comp} in cycle-bubbles. In both cases the algorithm starts on the dark blue edge of square 1. In (i) we have a balanced cycle-bubble, for which the resulting straight disambiguation and its complementary alternative have the same score (co-optimality). In (ii) we have an unbalanced cycle-bubble, for which the resulting straight disambiguation and its complementary alternative have distinct scores. }
    \label{fig:straight-algo}
\end{figure}

Two $\{6\}$-squares $\mathcal{Q}$ and $\mathcal{Q}'$ are \emph{neighbors} when a vertex of $\mathcal{Q}$ is connected to a vertex of $\mathcal{Q}'$ by a $\check{\mathbb{D}}$-edge.
Any $\mathbb{S}$-edge $e$ of a $\{6\}$-square $\mathcal{Q}$ in a cycle-bubble~$M$ is part of exactly one $\cpl$-cycle (Proposition~\ref{prop:Sedges}) and both $\check{\mathbb{D}}$-edges inciding at the endpoints of $e$ would clearly induce the same $\cpl$-cycle. For that reason, the choice of $e$ (and its paralogous edge~$\hat{e}$) implies a unique way of resolving all neighbors of $\mathcal{Q}$, and, by propagating this to the neighbors of the neighbors and so on, all squares of $M$ are resolved, resulting in what we call \emph{straight solution} $\tau_M$ (see Algorithms~\ref{alg:6-amb-comp} and~\ref{alg:6-rec-visit}). Then we can immediately obtain the \emph{complementary} alternative solution~$\widetilde{\tau}_M$, by switching all ambiguous squares of $\tau_M$. A cycle-bubble is said to be
\emph{unbalanced} if $\tau_M \ne \widetilde{\tau}_M$ or \emph{balanced} if $\tau_M = \widetilde{\tau}_M$. If $M$ is unbalanced, its score is given either by~$\tau_M$ or by~$\widetilde{\tau}_M$ (the maximum among the two). If $M$ is balanced, its score is given by both~$\tau_M$ and~$\widetilde{\tau}_M$ (co-optimality). Examples are given in Figure~\ref{fig:straight-algo}.

\begin{algorithm}[ht]
\caption{\label{alg:6-amb-comp}$\textsc{StraightBubbleSolution}$}
\begin{algorithmic}[1]
	\REQUIRE A cycle-bubble $M$ whose $\{6\}$-squares are numbered $\mathcal{Q}_1$, $\mathcal{Q}_2$, \ldots, $\mathcal{Q}_{m}$ 
	\ENSURE A solution $\tau_{M}$ of $M$
	\STATE $e \gets$ any $\mathbb{S}$-edge in $\mathcal{Q}_1$; 
	\STATE $\tau_M[1] \gets \{e, \hat{e}\}$;
	\STATE {\bf for} {~$i \gets 2,\ldots,m$~} {\bf do} ~$\tau_M[i] \gets \emptyset$;
	\STATE $\textsc{ResolveNeighbors}(\tau_M,e)$; \quad\COMMENT{recursive procedure}
	\IFTHEN{~$\hat{e}$ is an $\mathbb{S}$-edge in $M$~} \quad\COMMENT{the paralogous $\mathbb{S}$-edge $\hat{e}$ is also in $M$}
	\STATE ~~~$\textsc{ResolveNeighbors}(\tau_M,\hat{e})$; \quad\COMMENT{recursive procedure}
    \RETURN $\tau_M$
\end{algorithmic}
\end{algorithm}

\begin{algorithm}[ht]
\caption{\label{alg:6-rec-visit}$\textsc{ResolveNeighbors}$}%
\begin{algorithmic}[1]
	\REQUIRE A partially filled solution $\tau_M$ and an $\mathbb{S}$-edge $uv$ of cycle-bubble $M$
	
	\COMMENT{$\mathbb{S}$-edge $uv$ is adjacent to two $\check{\mathbb{D}}$-edges $uz$ and $vw$}
	\IFTHEN{~vertex $z$ is not in a resolved or fixed square~}
	\STATE ~~~$i \gets$ index in $\tau_M$ of square containing $z$;  
	\STATE ~~~$e \gets \mathbb{S}$-edge $zx$ of $\mathcal{Q}_i$ forming a $\cpl$-cycle with $uv$ and $uz$;
	\STATE ~~~$\tau_M[i] \gets \{e, \hat{e}\}$;
	\STATE ~~~{\bf if} ~$\hat{e}$ is an $\mathbb{S}$-edge in $M$~ {\bf then}
	\STATE ~~~~~~$\textsc{ResolveNeighbors}(\tau_M,\hat{e})$;
	
	\IFTHEN{~vertex $w$ is not in a resolved or fixed square~}
	\STATE ~~~$j \gets$ index in $\tau_M$ of square containing $w$;
	\STATE ~~~$f \gets \mathbb{S}$-edge $wy$ of $\mathcal{Q}_j$ forming a $\cpl$-cycle with $uv$ and $vw$;
	\STATE ~~~$\tau_M[j] \gets \{f, \hat{f}\}$;
	\STATE ~~~{\bf if} ~$\hat{f}$ is an $\mathbb{S}$-edge in $M$~ {\bf then}
	\STATE ~~~~~~$\textsc{ResolveNeighbors}(\tau_M,\hat{f})$;
    \RETURN
\end{algorithmic}
\end{algorithm}

\section{Solving the \boldmath$\sigma_6$ disambiguation with linear chromosomes}

\begin{figure}[ht!]
    \begin{center}
  \includegraphics[scale=0.36]{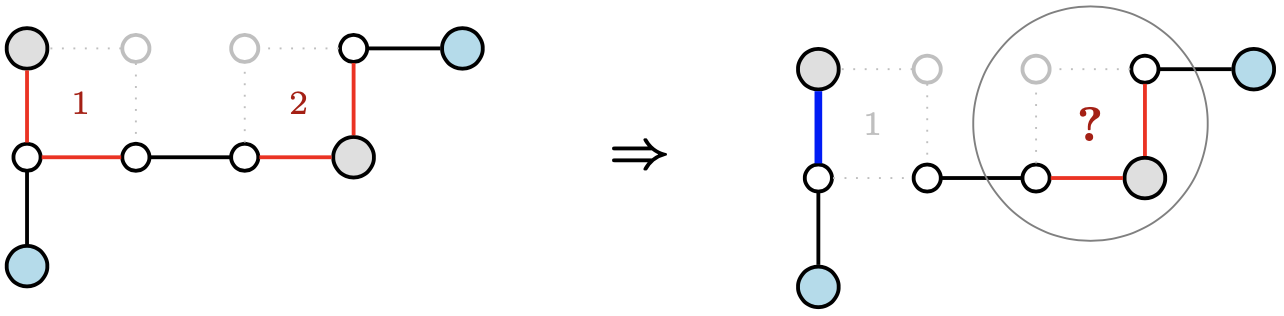}
  \end{center}
    
    \caption{Example showing that the straight algorithm does not work with paths: if we start on the dark blue edge of square number 1, we cannot propagate the effect of this choice to the neighbor square. }
    \label{fig:path-no-straight}
\end{figure}

For genomes with linear chromosomes, the ambiguous components might include paths besides cycle-bubbles. In the presence of paths, the straight algorithm unfortunately does not work (see Figure~\ref{fig:path-no-straight}). We must then proceed with an additional characterization of each ambiguous component~$M$ of $PG(\mathbb{S},\check{\mathbb{D}})$, splitting the disambiguation of~$M$ into smaller subproblems.

As we will present in the following, the solution for arbitrarily large components can be split into two types of problems, which are analogous to solving the \emph{maximal independent set} of auxiliary subgraphs that are either simple paths or double paths. In both cases, the solutions can be obtained in linear time.

\subsection{Intersection graph of an ambiguous component}

The auxiliary \emph{intersection graph} $\mathcal{I}(M)$ of an ambiguous component $M$ has a vertex with weight $\frac{1}{2}$ for each $\ppl$-path and a vertex with weight $1$ for each $\cpl$-cycle of $M$. Furthermore, if two distinct players intersect, we have an edge between the respective vertices. The intersection graphs of all ambiguous components 
can be built during the pruning procedure without increasing its linear time complexity. 

Note that an independent set of maximum weight in $\mathcal{I}(M)$ corresponds to an optimal solution of $M$. Although in general this problem is NP-hard, 
in our case the underlying ambiguous component $M$ imposes a regular structure to its intersection graph, allowing us to find such an independent set in linear time.

If two $\ppl$-paths intersect in their $\g{S}$-telomere, this intersection must include the incident $\check{\g{D}}$-edge.
Therefore, when we say that an intersection occurs at an $\g{S}$-telomere, this automatically means that the intersection is the $\check{\g{D}}$-edge inciding in an $\g{S}$-telomere.
A valid 4-cycle has two $\check{\g{D}}$-edges and a valid 6-cycle has three $\check{\g{D}}$-edges. Besides the one at the $\g{S}$-telomere, a valid 4-path has one $\check{\g{D}}$-edge while a valid 2-path has none - therefore the latter cannot intersect with a $\cpl$-cycle.
When we say that 4-paths and/or $\cpl$-cycles intersect with each other in a $\check{\g{D}}$-edge, we refer to an  \emph{inner} $\check{\g{D}}$-edge not one inciding in an $\g{S}$-telomere.

Since the contribution of each cycle in the score is twice as much as the contribution of a path, we make a distinction between two types of subgraphs of an intersection graph $\mathcal{I}(M)$, which can correspond to cycle-bubbles or path-flows.

\subsection{Path-flows in the intersection graph}

\begin{figure}[ht]
    \begin{center}

  \includegraphics[scale=0.36]{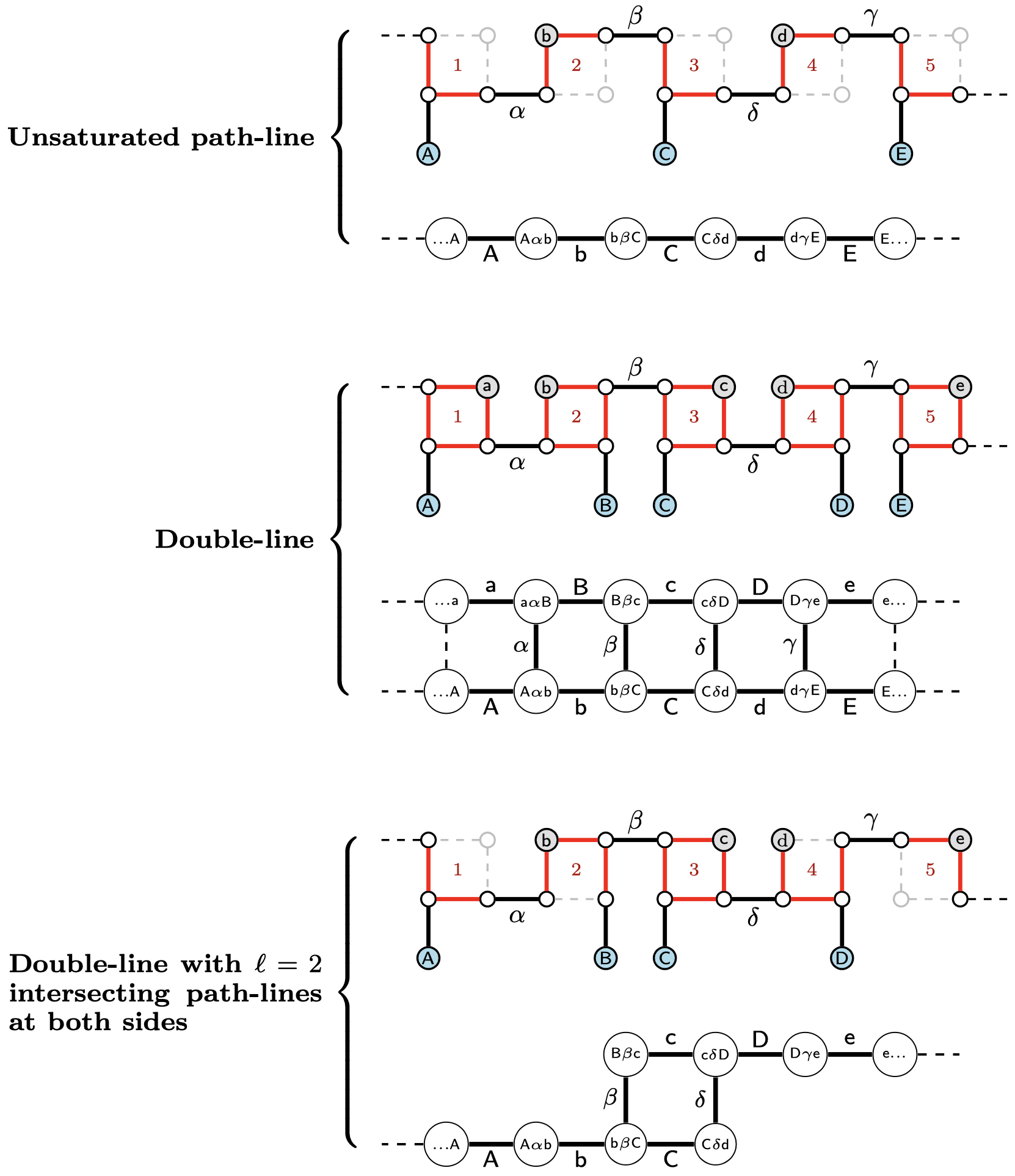}

    \end{center}

	 \caption{Examples of an unsaturated path-line, a double-line and the intersection between a double-line and two unsaturated path-lines.
	}
    \label{fig:path-lines}
\end{figure}

A \emph{path-flow} in $\mathcal{I}(M)$ is a maximal connected subgraph whose vertices correspond to $\ppl$-paths. 
A \emph{path-line} of length $\ell$ in a path-flow is a series of $\ell$ paths, such that each pair of  consecutive paths intersect at a telomere. Assume that the vertices in a path-line are numbered from left to right with integers $1,2,\ldots,\ell$.
A \emph{double-line} consists of two parallel path-lines of the same length $\ell$, such that vertices with the same number in both lines intersect in a $\check{\g{D}}$-edge and are therefore connected by an edge. A 2-path has no free $\check{\g{D}}$-edge, therefore a double-line is exclusively composed of 4-paths. 
If a path-line composes a double-line, it is \emph{saturated}, otherwise it is \emph{unsaturated}.
Since each 4-path of a double-line has a $\check{\g{D}}$-edge intersection with another and each 4-path can have only one $\check{\g{D}}$-edge intersection, no vertex of a double-line can be connected to a cycle in $\mathcal{I}(M)$.
Examples of an unsaturated path-line and a double-line are given in Figure~\ref{fig:path-lines}.

\begin{figure}
    \begin{center}
  \includegraphics[scale=0.36]{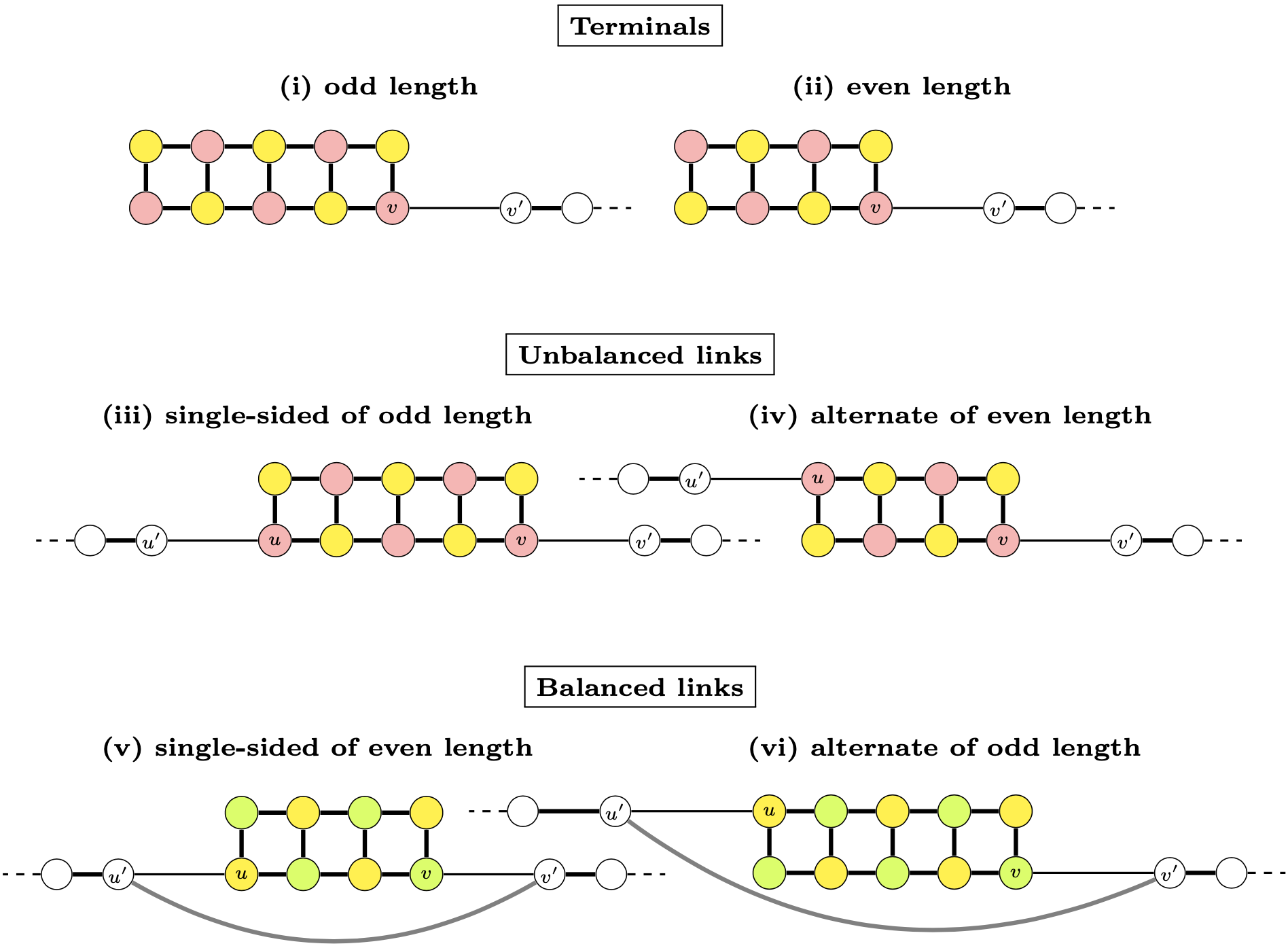}
  \end{center}
    
	 \caption{Types of double-line: terminal, balanced and unbalanced links. The yellow solution that in cases (i-ii) leaves $v$ unselected and in cases (iii-iv) leaves $u$ and $v$ unselected can be fixed so that an independent set of the adjacent unsaturated path-line(s) can start at $v'$ (and $u'$). In cases (v-vi) either the yellow or the green solution will be fixed later; it will be the one compatible with the selected independent set of the unsaturated path-line ending in $u'$ concatenated to the one starting in $v'$. }
    \label{fig:double-line-types}
\end{figure}

Let us assume that a double-line is always represented with one upper path-line and one lower path-line.
A double-line of length $\ell$ has $2\ell$ vertices and exactly two independent sets of maximal weight, each one with  $\ell$ vertices and weight $\frac{\ell}{2}$: one includes the paths with odd numbers in the upper line and the paths with even numbers in the lower line, while the other includes the paths with even numbers in the upper line and the paths with odd numbers in the lower line. Since a double-line cannot intersect with cycles, it is clear that at least one of these independent sets will be part of a global optimal solution for~$\mathcal{I}(M)$. In other words, not only the two possible local optimal solutions and their (common) weight are known, but it is guaranteed that at least one of them will be part of a global optimal solution. 
A maximal double-line can be of three different types:

\begin{enumerate}
    \item \emph{Isolated}: corresponds to the complete graph $\mathcal{I}(M)$. Here 
    the double line can be cyclic. If $\ell$ is even, in both upper and lower lines of a cyclic double-line, the last vertex intersects at a telomere with the first vertex. If $\ell$ is odd, this connection of a cyclic double-line is ``twisted'': the last vertex of the upper line intersects at a telomere with the first vertex of the lower line, and the first vertex of the upper line intersects at a telomere with the last vertex of the lower line. 
    Being cyclic or not, any of the two optimal local solutions can be fixed.
    \item \emph{Terminal}: intersects with one unsaturated path-line, and, without loss of generality, the intersection involves the vertex $v$ located at the rightmost end of the lower line. Here at least one of the two optimal local solutions would leave $v$ unselected; we can safely fix this option. (See Figure~\ref{fig:double-line-types}.)

    \item \emph{Link}: intersects with unsaturated lines at both ends. The\;intersections\;can\;be:
    \begin{enumerate}
        \item \emph{single-sided}: both occur at the ends of the same saturated line, or 
        \item \emph{alternate}: the left intersection occurs at the end of one saturated line and the right intersection occurs at the end of the other. 
    \end{enumerate}
    \smallskip
    Let $v'$ be the outer vertex connected to a vertex $v$ belonging to the link at the right and $u'$ be the outer vertex connected to a vertex $u$ belonging to the link at the left.
Let a \emph{balanced link} be alternate of odd length, or single-sided of even length.
In contrast, an \emph{unbalanced link} is alternate of even length, or single-sided of odd length.
If the link is unbalanced, one of the two local optimal solutions leaves both~$u$ and $v$ unselected; we can safely fix this option.
If the link is balanced, we cannot fix the solution before-hand, but we can reduce the problem, by removing the connections $uu'$ and $vv'$ and adding the connection $u'v'$.
Since both $u'$ and~$v'$ must be the ends of unsaturated lines, this procedure simply concatenates these two lines into a single unsaturated path-line.
(See Figure~\ref{fig:double-line-types}~(v) and~(vi).)
Finding a maximum independent set of the remaining unsaturated path-lines is a trivial problem that will be solved last; depending on whether one of the vertices $u'$ and $v'$ is selected in the end, we can fix the solution of the original balanced link.
\end{enumerate}

\subsection{Intersection between path-flows and cycle-bubbles}

If an ambiguous component has only cycles, its solution can be easily obtained with the straight algorithm presented in the previous section. More intricate is when an ambiguous component $M$ includes cycles and paths. In this case we redefine a \emph{cycle-bubble} as corresponding to a maximal connected subgraph of $\mathcal{I}(M)$ whose vertices correspond to $\cpl$-cycles. Let $H$ be the subgraph of $M$ including all edges that compose the cycles of a cycle-bubble. An optimal solution 
for $H$ is either the straight solution $\tau_H$, given by Algorithm~\ref{alg:6-amb-comp}, or its alternative~$\widetilde{\tau}_H$. Recall that if both~$\tau_H$ and~$\widetilde{\tau}_H$ have the same score, then $H$ is said to be \emph{balanced}, otherwise it is said to be \emph{unbalanced}. 

\begin{proposition}\label{prop:cycles-dominate}
   Let an ambiguous component $M$ have cycle-bubbles $H_1$, ..., $H_q$. There is an optimal solution for~$M$ including, for each $i=1,...,q$: (1) the optimal solution for~$H_i$, if~$H_i$ is unbalanced; or (2) either~$\tau_{H_i}$ or~$\widetilde{\tau}_{H_i}$, if~$H_i$ is balanced.
\end{proposition} 

\begin{figure}
    \begin{center}
  \includegraphics[scale=0.36]{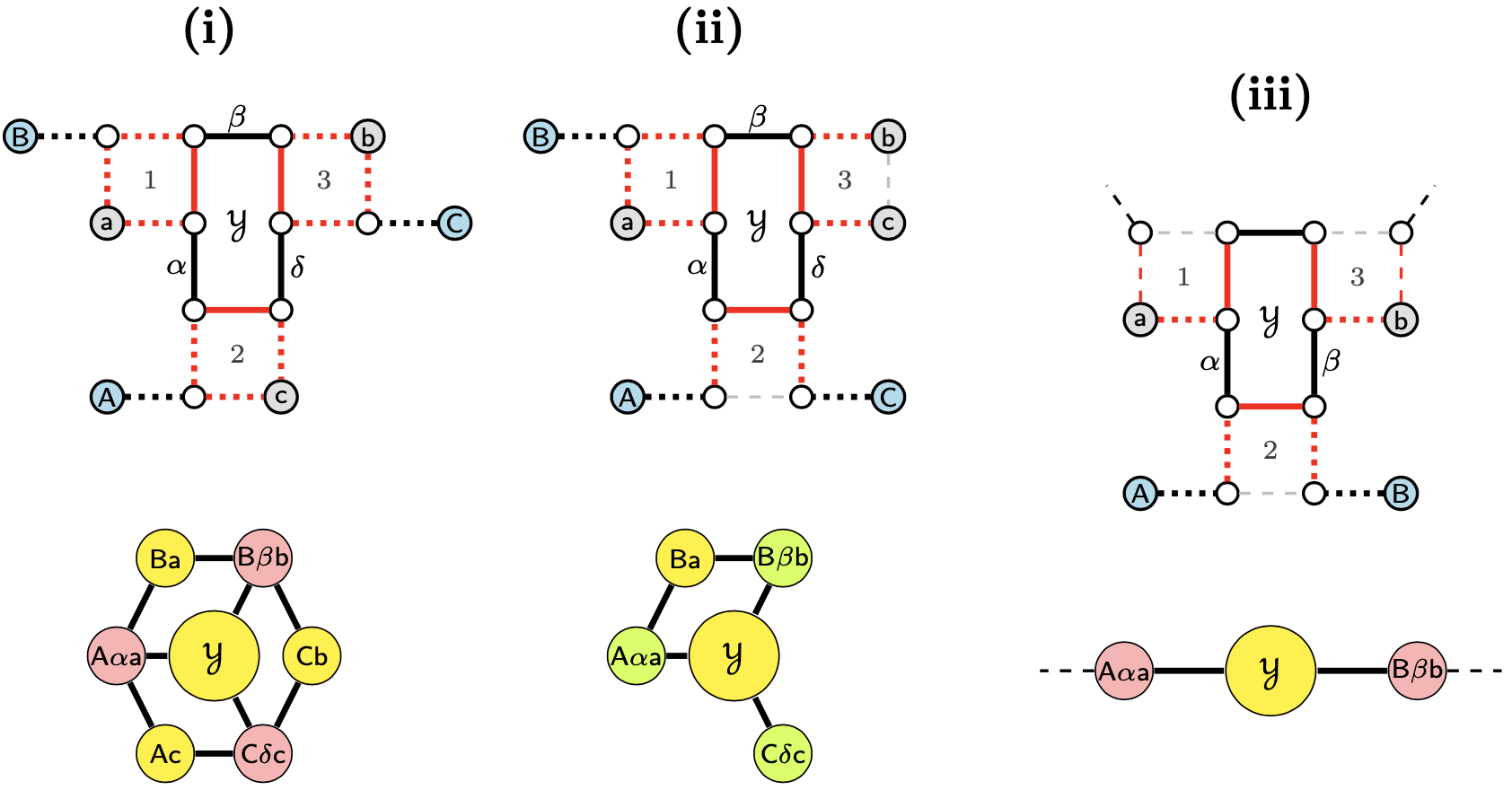}
  \end{center}
    
   \caption{Underlying pruned subgraphs and corresponding intersection graphs of a bubble with a single 6-cycle $\mathcal{Y}$ (solid edges). Dotted edges are exclusive to paths and dashed gray edges are pruned out.
    In (i) and (ii), $\mathcal{Y}$ intersects with three valid 4-paths $\B{A}\alpha\B{a}$, $\B{B}\beta\B{b}$ and $\B{C}\delta\B{c}$. In (i), the yellow solution including $\mathcal{Y}$ would also include the three 2-paths $\B{Ab}$, $\B{Bc}$ and $\B{Ca}$, being clearly superior. In (ii), the yellow solution including $\mathcal{Y}$ would still include the 2-path $\B{Ba}$, having the same score of the green solution with three 4-paths. In any of the two cases, the underlying graph cannot be extended. In (iii), $\mathcal{Y}$ has plug connections with unsaturated path-lines starting at 4-paths $\B{A}\alpha\B{a}$ and $\B{B}\beta\B{b}$ (both can be extended).
    }
    \label{fig:cycles-dominate-1}
\end{figure}

\proof 
We will analyze the cases by increasing the size of the maximal subgraph containing intersecting cycles:
\begin{enumerate}
    \item \textit{A $\{4,6\}$-cycle $C$ that does not intersect with any other $\{4,6\}$-cycle}: (a) if $C$ is a 4-cycle, it can intersect with at most two valid 4-paths; therefore there is an optimal solution including $C$; (b) if $C$ is a 6-cycle, it can intersect with at most three valid 4-paths, but if it intersects with three valid 4-paths there will be at least one valid 2-path $P$ compatible with $C$ ; therefore there is an optimal solution including $C$ and $P$ (see Figure~\ref{fig:cycles-dominate-1}~(ii)).
    
    \item \textit{Two $\{4,6\}$-cycles $C$ and $C'$ intersecting with each other but not with any other $\{4,6\}$-cycle}: Since valid 4-cycles have less edges for intersection, let us assume without loss of generality that both $C$ and $C'$ are 6-cycles. Their intersection (illustrated in Figures~\ref{afig:int2bub-sat} and~\ref{afig:int2bub-unsat} of Appendix~\ref{app:not-plugs}) can be:
    \begin{itemize}
    \item[(a)] a $\check{\g{D}}\g{S}\check{\g{D}}$-path, and in this case each cycle can intersect with at most one valid 4-path, therefore there is an optimal solution including either $C$ or~$C'$;
    \item[(b)] a single $\check{\g{D}}$-edge, and in this case each cycle can intersect with two valid 4-paths, therefore there is an optimal solution including either $C$ or~$C'$.
    \end{itemize}
  \end{enumerate}
  As the size of the bubble grows, there is less space for intersecting paths, and each cycle intersects with at most one path.
  In general, the best we can get by replacing cycles by paths are  co-optimal solutions.
\qed

As a consequence of Proposition~\ref{prop:cycles-dominate}, if a cycle-bubble is unbalanced, its optimal solution can be fixed so that the unsaturated path-lines around it can be treated separately. Similarly, if a balanced cycle-bubble $H$ has a single intersection involving a cycle $C$ and a path $P$ (that can be the first vertex of an unsaturated path-line), then we can immediately fix the solution of $H$ that does not contain $C$.

\subsubsection{Balanced cycle-bubbles intersecting with at least two paths.} 
\begin{figure}[ht]
  \begin{center}
  \includegraphics[scale=0.36]{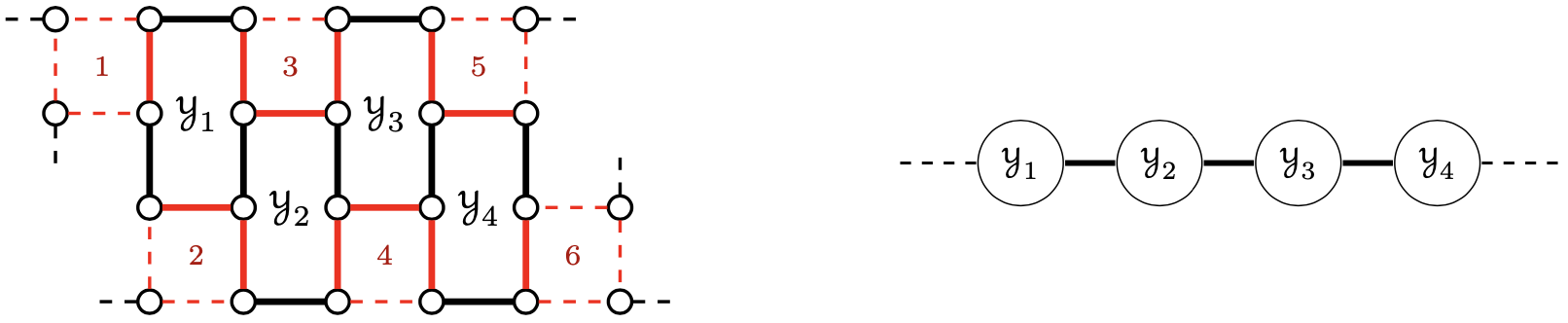}
  \end{center}
 	
	\caption{Cycle-bubble of type cycle-line and its intersection graph.}
    \label{fig:cycle-line}
\end{figure}

If a cycle-bubble $H$ is balanced and intersects with at least two paths, then it requires a special treatment. 
However, as we will see, here the only case that can be arbitrarily large is easy to handle. Let a cycle-bubble be a \emph{cycle-line} when it consists of a series of valid 6-cycles, such that each pair of consecutive cycles intersect at a $\check{\g{D}}$-edge (see Figure~\ref{fig:cycle-line}).

\begin{proposition}\label{prop:max-bubble-size}
  Cycle-bubbles involving 9 or more cycles must be a cycle-line.
\end{proposition}

\proof
In Figure~\ref{fig:complex-bubble} (whose steps are more elaborated in Figures~\ref{afig:2bub}-\ref{afig:8bub} of Appendix~\ref{app:complex}) we show that, if a bubble is not a line, it reaches its ``capacity'' with at most 8 cycles.
\qed

\begin{figure}[ht!]
    \begin{center}
  \includegraphics[scale=0.36]{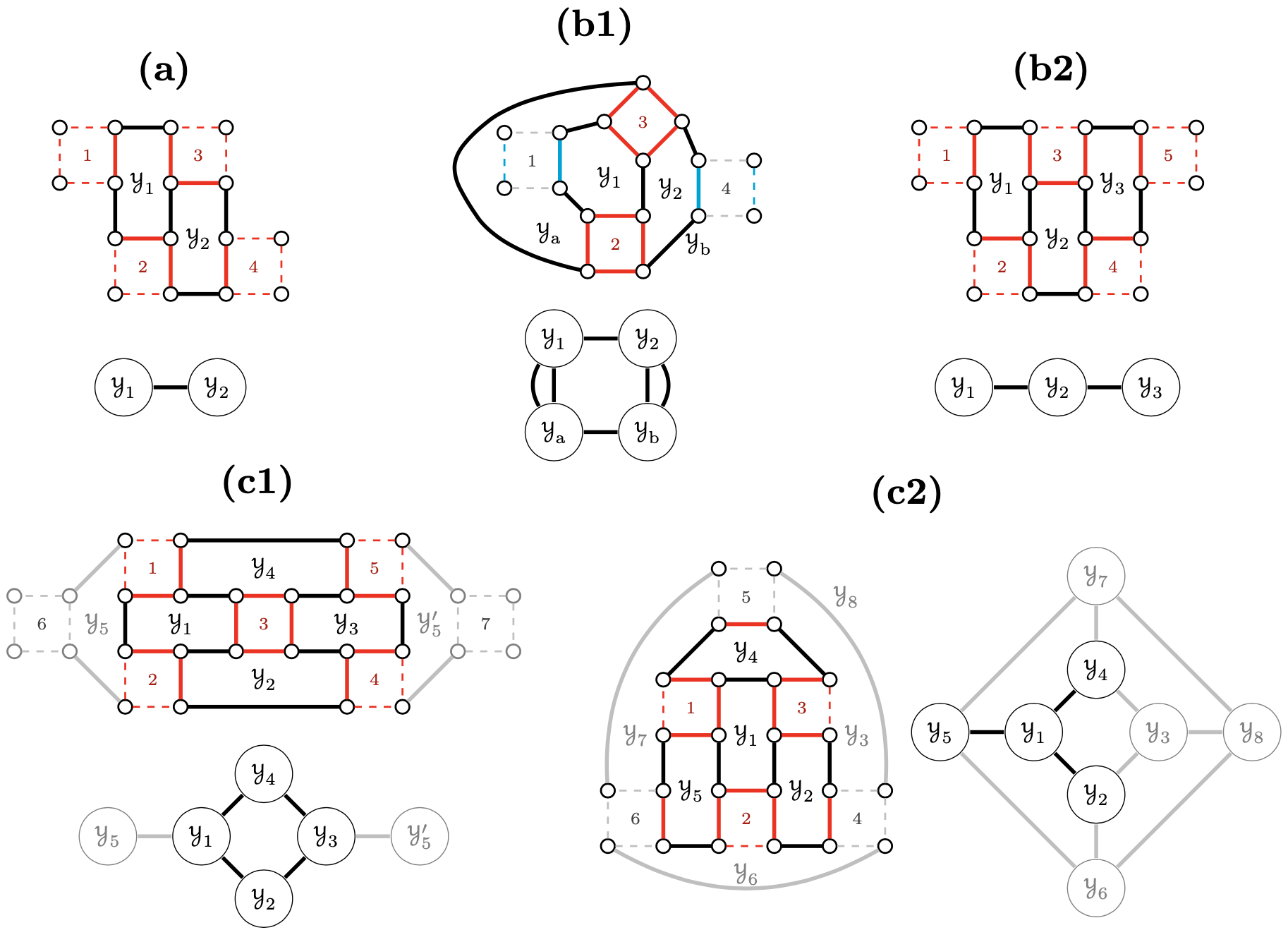}
  \end{center}
 
    \caption{While a cycle-line can be arbitrarily large, by increasing the complexity of a bubble we quickly saturate the space for adding cycles to it. 
    Starting with (a) a simple cycle-line of length two, we can either (b1) connect the open vertices of squares 2 and 3, obtaining a cyclic cycle-line of length 4 that cannot be extended, or (b2) extend the line so that it achieves length three. From (b2) we can obtain (c1) a cyclic cycle-line of length 4 that can be extended first by adding cycle $\mathcal{Y}_5$ next to $\mathcal{Y}_1$ and then either adding $\mathcal{Y}_5'$ next to $\mathcal{Y}_3$ or closing $\mathcal{Y}_6$, $\mathcal{Y}_7$ and $\mathcal{Y}_8$ so that we get (c2). In both cases no further extensions are possible. Note that (c2) can also be obtained by extending a cycle-line of length three and transforming it in a \emph{star} with three branches, that can still be extended by closing $\mathcal{Y}_3$, $\mathcal{Y}_6$, $\mathcal{Y}_7$ and $\mathcal{Y}_8$. (These steps are more elaborated in Figures~\ref{afig:2bub}-\ref{afig:8bub} of Appendix~\ref{app:complex}.)}
    \label{fig:complex-bubble}
\end{figure}

\smallskip

Besides having its size limited to 8 cycles, the more complex a non-linear cycle-bubble becomes, the less space it has for paths around it. The solutions for these few exceptional bounded cases are described in the end of this section.

\smallskip

Our focus now is the remaining situation of a balanced cycle-line with intersections involving at least two cycles. 
Recall that cycles can only intersect with unsaturated path-lines. An intersection between a cycle- and a path-line is a \emph{plug connection} when it occurs between vertices that are at the ends of both lines.

\begin{proposition}\label{prop:plug-intersections}
  Cycle-lines of length at least 4 can only have plug connections. 
\end{proposition}

\begin{figure}
     \begin{center}
  \includegraphics[scale=0.36]{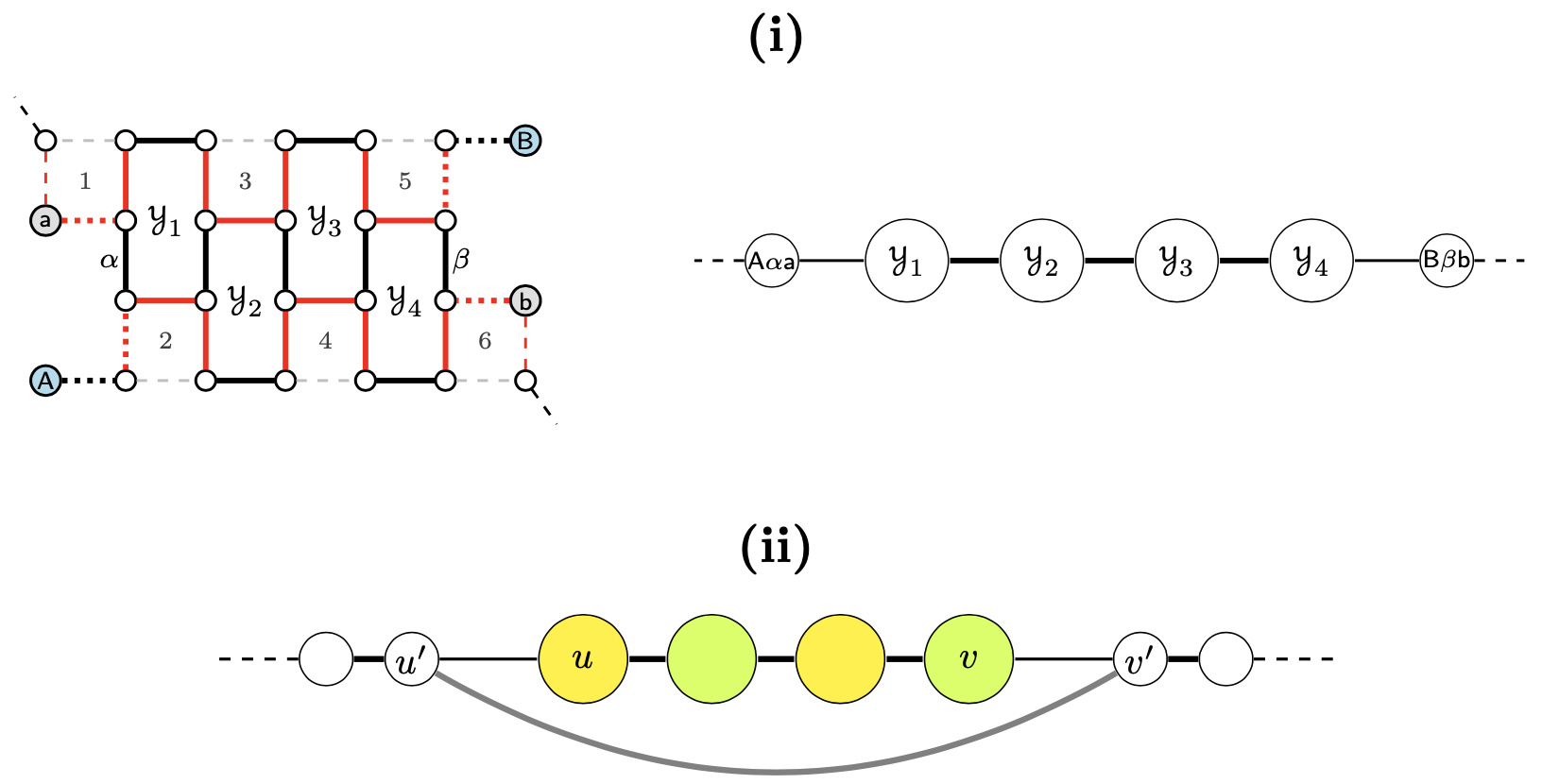}
  \end{center}

    \caption{(i) A cycle-line of length 4 or larger only allows plug connections. In contrast, cycle-lines of lengths 1-3 admit other types of connections (see Figures~\ref{fig:cycles-dominate-1} and~\ref{afig:int2bub-sat}-\ref{afig:int2bub-unsat}, the last two in Appendix~\ref{app:not-plugs}). (ii) If the cycle-line has even length and plug connections at both sides, we have a balanced link: either the yellow or the green solution will be fixed later; it will be the one compatible with the selected independent set of the unsaturated path-line ending in $u'$ concatenated to the one starting in $v'$.}
    \label{fig:cycle-line-4}
\end{figure}

\proof
If a cycle-line has length at least four, its underlying graph has only  ``room'' for intersections with 4-paths next to its leftmost of rightmost cycles. See the illustration in Figure~\ref{fig:cycle-line-4}.
\qed

\smallskip
For arbitrarily large instances, the last missing case is of a balanced cycle-line with plug connections at both sides, called a \emph{balanced link}. The procedure here is the same as that for double-lines that are balanced links, where the local solution can only be fixed after fixing those of the outer connections (see Figure~\ref{fig:cycle-line-4}~(ii)).

\paragraph{Exceptional bounded cases.} 

Balanced cycle-lines with two cycles can have connections to path-lines that are not plugs, but the number of cases is again limited. In most of them (shown in Figure~\ref{afig:int2bub-sat} of Appendix~\ref{app:not-plugs}) the bubble is saturated and the paths around cannot be connected to extendable path-lines. 
For these bubbles all paths are over the same squares of the cycles, therefore the  straight algorithm would give the two overall alternatives including the paths around each of these bubbles, and the best solution can be immediately fixed.

In another case (shown in Figure~\ref{afig:int2bub-unsat}~(i) of Appendix~\ref{app:not-plugs}) there is one extendable path-line, but the local solution (including the bubble and the paths that are over the same squares) is unbalanced, therefore also here we can fix the best among the two overall alternatives given by the straight algorithm.

In the last two cases (shown in Figure~\ref{afig:int2bub-unsat}~(ii)~and~(iii) of Appendix~\ref{app:not-plugs}) there are extendable path-lines, and the local solutions (including the bubble and the paths that are over the same squares) are balanced. In the first case, there is only one extendable path-line and we can fix the solution including the cycle that is connected to last ``visible'' path of the path-line. The second case is analogous to cycle-lines of type balanced link, with the difference that here the lines are already concatenated; the local solution can then only be fixed after fixing those of the outer connections.

Concerning non-linear cycle-bubbles, there are only four distinct cases that need to be considered: one case of a non-linear bubble with two 6-cycles (Figure~\ref{afig:int2bub-sat}~(iii) of Appendix~\ref{app:not-plugs}) and three cases of non-linear bubbles with four 6-cycles (Figure~\ref{afig:int4bub-sat} in Appendix~\ref{app:not-plugs}). In all of these four cases, the bubble is saturated and the paths around cannot be connected to extendable path-lines. Indeed, also for these bubbles all paths are over the same squares of the cycles, therefore the  straight algorithm would give the two overall alternatives including the paths around each of these bubbles, and the best among these solutions can be immediately fixed.

\subsubsection{What remains is a set of independent unsaturated path-lines.} 
If what remains is a single unsaturated path-line of even length, it can even be cyclic\footnote{Indeed, a cyclic unsaturated line of two paths is the exceptional case in which they intersect at both telomeres (see Figure~\ref{afig:int2bub-sat}~(iii) in Appendix~\ref{app:not-plugs}).}.
In any case, an optimal solution 
can be trivially found. First assume that in an unsaturated path-line of length $\ell$ the paths are numbered from left to right with $1,2,\ldots,\ell$. The solution that selects all paths with odd numbers must be optimal. Fix this solution and, depending on the connections between the selected 
vertices of the unsaturated path-line and vertices from  balanced links that are double-lines or cycle-lines, fix the compatible solutions for the latter ones.

\section{\label{sec:discussion}Final remarks and discussion}

Given a singular genome $\g{S}$ and a duplicated genome $\g{D}$ over the same set of gene families, the double distance of~$\g{S}$ and~$\g{D}$ aims to find the smallest distance between $\g{D}$ and any element from the set $\T{2}\g{S}$, that contains all possible genome configurations obtained by doubling the chromosomes of $\g{S}$.
Different underlying genomic distance measures give rise to different double distances: the breakpoint double distance of $\g{S}$ and $\g{D}$ is an easy problem that can be greedily solved in linear time, while computing the DCJ double distance of $\g{S}$ and $\g{D}$ is NP-hard.
Our study is an exploration of the complexity space between these two extremes.

We considered a class of genomic distance measures called $\sigma_k$ distances, for $k=2,4,6,\ldots,\infty$, which are between the breakpoint ($\sigma_2$) and the DCJ ($\sigma_\infty$) distance.
In this work we presented linear time algorithms for computing the double distance under the~$\sigma_4$, and under the $\sigma_6$ distance. Our solution relies on a variation of the breakpoint graph called ambiguous breakpoint graph.

The solutions we found so far are greedy with all players being optimal in $\sigma_2$, greedy with all players being co-optimal in $\sigma_4$ and non-greedy with non-optimal players in $\sigma_6$, all of them running in linear time.
More specifically for the $\sigma_6$ case, after a pre-processing that fixes symmetric squares and triplets, at most two players share an edge. However we can already observe that, as $k$ grows, the number of players sharing a same edge also grows. For that reason, we believe that, if for some $k\geq8$ the complexity of the $\sigma_k$ double distance is found to be NP-hard, the complexity is also NP-hard for any~$k'>k$. 
We expect that when we find the smallest $k$ for which the $\sigma_k$ double distance is NP-hard we will be able to confirm this conjecture.
In any case, the natural next step in our research is to study the $\sigma_8$ double distance. 

Besides the double distance, other combinatorial problems related to genome evolution and ancestral reconstruction, including median and guided halving, have the distance problem as a basic unit. And, analogously to the double distance, these problems can be solved in polynomial time (but differently from the double distance, not greedy and linear) when they are built upon the breakpoint distance, while they are NP-hard when they are built upon the DCJ distance~\cite{TAN-ZHE-SAN-2009}. Therefore, a challenging avenue of research is doing the same exploration for both median and guided halving problems under the class of $\sigma_k$ distances. 
In both cases it seems possible to adopt variations of the breakpoint graph. 
To the best of our knowledge, the guided halving problem has not yet been studied for any $\sigma_k$ distance except $k=2$ and $k=\infty$, while for the median much effort for the $\sigma_4$ distance has been done but no progress was obtained so far. A reason for this difference of progress between double distance and median is probably related to the underlying approaches. While the double distance can be solved by \emph{removing} paralogous edges from the ambiguous breakpoint graph, solving the median requires \emph{adding} new edges (representing the adjacencies of the median genome) to an extended (\emph{multiple}) breakpoint graph, and the combinatorial space of the distinct possibilities of doing that could not yet be described.

\section*{Acknowledgements}
We would like to thank Cedric Chauve for bringing our attention to the class of $\sigma_k$ distances as a means for studying the hardness bound between the breakpoint distance and the DCJ distance in combinatorial problems related to genome evolution. Thanks also to Eloi Araujo, Daniel Doerr and Fábio H. V. Martinez for helping us studying the median problem under this class.

\clearpage


\bibliographystyle{splncs04} 
\bibliography{bibliography}      

\clearpage
\appendix

\medskip
\section{Supplementary figures}\label{app:1}

\newcounter{myfigure}[section]
\refstepcounter{myfigure}

\renewcommand\thefigure{\thesection\arabic{myfigure}} 

\medskip
All the figures presented here assume a graph free of symmetric squares and triplets. For each case we have the ambiguous component of the pruned graph and its intersection graph. Often small modifications (e.g., by switching the positions of $\g{S}$- and $\check{\g{D}}$-telomeres) lead to equivalent cases, and here we show only one of these. In the particular situations of an intersection between two cycles being a $\check{\g{D}}\g{S}\check{\g{D}}$-path or  intersections between two paths occurring at both telomeres, the respective vertices of the intersection graph are connected by two parallel edges.

\medskip
\subsection{Complex bubbles are limited to 8 cycles}\label{app:complex}

\smallskip
By a complete enumeration of cases, in Figures~\ref{afig:2bub}-\ref{afig:8bub} we show that, if a bubble is not a line, it reaches its ``capacity'' with at most 8 cycles.
In all figures dashed gray edges are pruned out.

\smallskip
\input{appendix/Supplementary-Figures/bubbles}

\clearpage

\subsection{Balanced cycle-bubbles intersecting with more than one path}\label{app:not-plugs}

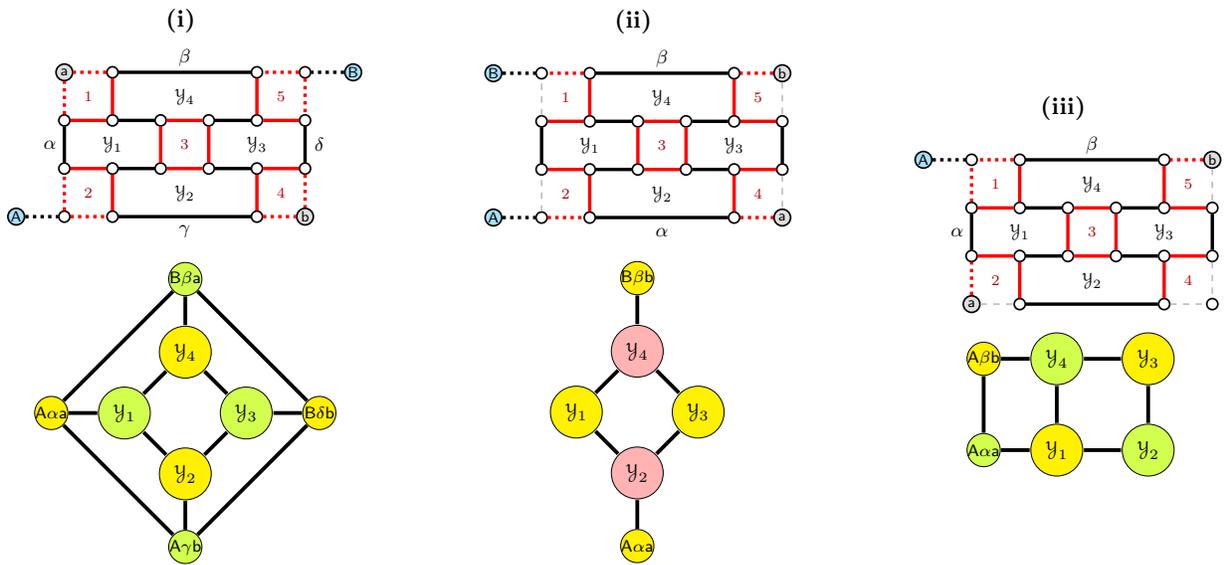
\begin{figure}[ht!]
	\begin{center}

	\begin{minipage}{5cm} 
    \begin{center}
     \textbf{(i)}
    \smallskip
    
    \scalebox{0.8}{
		\begin{tikzpicture}[scale=0.8]
			
			\vcl{(0,2)}{u1}{}{white}
			\tvcl{(0,3)}{v1}{$\B{a}$}{telD}
			\vcl{(1,3)}{uu1}{}{white}
			\vcl{(1,2)}{vv1}{}{white}

			\node at (0.5,2.5) {\color{darkred}\scriptsize 1};
			\node at (0.5,0.5) {\color{darkred}\scriptsize 2};
			\node at (2.5,1.5) {\color{darkred}\scriptsize 3};
			\node at (4.5,0.5) {\color{darkred}\scriptsize 4};
			\node at (4.5,2.5) {\color{darkred}\scriptsize 5};
			
			\node at (1,1.5) {$\mathcal{Y}_1$};
			\node at (2.5,0.5) {$\mathcal{Y}_2$};
			\node at (4,1.5) {$\mathcal{Y}_3$};
			\node at (2.5,2.5) {$\mathcal{Y}_4$};
						
			\dted{u1}{v1}{red}
			\ed{u1}{vv1}{red}
			\dted{uu1}{v1}{red}
			\ed{uu1}{vv1}{red}
			
			\vcl{(0,0)}{u2}{}{white}
			\vcl{(0,1)}{v2}{}{white}
			\vcl{(1,1)}{uu2}{}{white}
			\vcl{(1,0)}{vv2}{}{white}
			
			\dted{u2}{v2}{red}
			\dted{u2}{vv2}{red}
			\ed{uu2}{v2}{red}
			\ed{uu2}{vv2}{red}
			
			\vcl{(2,1)}{u3}{}{white}
			\vcl{(2,2)}{v3}{}{white}
			\vcl{(3,2)}{uu3}{}{white}
			\vcl{(3,1)}{vv3}{}{white}
			
			\ed{u3}{v3}{red}
			\ed{u3}{vv3}{red}
			\ed{uu3}{v3}{red}
			\ed{uu3}{vv3}{red}
			
			\ed{vv1}{v3}{black}
			\ed{u3}{uu2}{black}
			\ed{u1}{v2}{black}
			
			\vcl{(4,0)}{u4}{}{white}
			\vcl{(4,1)}{v4}{}{white}
			\vcl{(5,1)}{uu4}{}{white}
			\tvcl{(5,0)}{vv4}{$\B{b}$}{telD}
			
			\ed{u4}{v4}{red}
			\dted{u4}{vv4}{red}
			\ed{uu4}{v4}{red}
			\dted{uu4}{vv4}{red}
			
			\vcl{(4,2)}{u5}{}{white}
			\vcl{(4,3)}{v5}{}{white}
			\vcl{(5,3)}{uu5}{}{white}
			\vcl{(5,2)}{vv5}{}{white}
			
			\ed{u5}{v5}{red}
			\ed{u5}{vv5}{red}
			\dted{uu5}{v5}{red}
			\dted{uu5}{vv5}{red}
			
			\ed{u4}{vv2}{black}
			\ed{v4}{vv3}{black}
			\ed{u5}{uu3}{black}			
			
			\ed{v5}{uu1}{black}
			\ed{vv5}{uu4}{black}
			
			\tvcl{(-1,0)}{t1}{$\B{A}$}{telS}
			\tvcl{(6,3)}{t2}{$\B{B}$}{telS}
			
			\dted{t1}{u2}{black}
			\dted{t2}{uu5}{black}
			
			\node at (-0.3,1.5) {$\alpha$};
			\node at (2.5,3.3) {$\beta$};
			\node at (5.3,1.5) {$\delta$};
			\node at (2.5,-0.3) {$\gamma$};
			
		\end{tikzpicture}
	}
	
	\medskip
	\scalebox{0.9}{
		\begin{tikzpicture}[scale=0.9]
			
			\pvcl{(-1.2,-1)}{c5}{\scriptsize $\B{A}\alpha\B{a}$}{yellow}
			\cvcl{(0,-1)}{c1}{$\mathcal{Y}_1$}{mylime}
			\cvcl{(1,0)}{c4}{$\mathcal{Y}_4$}{yellow}
			\cvcl{(1,-2)}{c2}{$\mathcal{Y}_2$}{yellow}
			\cvcl{(2,-1)}{c3}{$\mathcal{Y}_3$}{mylime}
			\pvcl{(1,1.2)}{c7}{\scriptsize $\B{B}\beta\B{a}$}{mylime}
			\pvcl{(1,-3.2)}{c6}{\scriptsize $\B{A}\gamma\B{b}$}{mylime}
			\pvcl{(3.2,-1)}{c8}{\scriptsize $\B{B}\delta\B{b}$}{yellow}
			\ed{c1}{c2}{black}
			\ed{c2}{c3}{black}
			\ed{c3}{c4}{black}
			\ed{c4}{c1}{black}
			\ed{c5}{c1}{black}
			\ed{c7}{c4}{black}
			\ed{c7}{c5}{black}
			\ed{c6}{c5}{black}
			\ed{c6}{c2}{black}
			\ed{c3}{c8}{black}
			\ed{c7}{c8}{black}
			\ed{c6}{c8}{black}

		\end{tikzpicture}
    }
    
	\end{center}
	\end{minipage}
	\hspace{8mm}
	\begin{minipage}{5cm} 
    \begin{center}
     \textbf{(ii)}
    \smallskip
    
    \scalebox{0.8}{
		\begin{tikzpicture}[scale=0.8]
			
			\vcl{(0,2)}{u1}{}{white}
			\vcl{(0,3)}{v1}{}{white}
			\vcl{(1,3)}{uu1}{}{white}
			\vcl{(1,2)}{vv1}{}{white}

			\node at (0.5,2.5) {\color{darkred}\scriptsize 1};
			\node at (0.5,0.5) {\color{darkred}\scriptsize 2};
			\node at (2.5,1.5) {\color{darkred}\scriptsize 3};
			\node at (4.5,0.5) {\color{darkred}\scriptsize 4};
			\node at (4.5,2.5) {\color{darkred}\scriptsize 5};
			
			\node at (1,1.5) {$\mathcal{Y}_1$};
			\node at (2.5,0.5) {$\mathcal{Y}_2$};
			\node at (4,1.5) {$\mathcal{Y}_3$};
			\node at (2.5,2.5) {$\mathcal{Y}_4$};
						
			\ded{u1}{v1}{lightgray}
			\ed{u1}{vv1}{red}
			\dted{uu1}{v1}{red}
			\ed{uu1}{vv1}{red}
			
			\vcl{(0,0)}{u2}{}{white}
			\vcl{(0,1)}{v2}{}{white}
			\vcl{(1,1)}{uu2}{}{white}
			\vcl{(1,0)}{vv2}{}{white}
			
			\ded{u2}{v2}{lightgray}
			\dted{u2}{vv2}{red}
			\ed{uu2}{v2}{red}
			\ed{uu2}{vv2}{red}
			
			\vcl{(2,1)}{u3}{}{white}
			\vcl{(2,2)}{v3}{}{white}
			\vcl{(3,2)}{uu3}{}{white}
			\vcl{(3,1)}{vv3}{}{white}
			
			\ed{u3}{v3}{red}
			\ed{u3}{vv3}{red}
			\ed{uu3}{v3}{red}
			\ed{uu3}{vv3}{red}
			
			\ed{vv1}{v3}{black}
			\ed{u3}{uu2}{black}
			\ed{u1}{v2}{black}
			
			\vcl{(4,0)}{u4}{}{white}
			\vcl{(4,1)}{v4}{}{white}
			\vcl{(5,1)}{uu4}{}{white}
			\tvcl{(5,0)}{vv4}{$\B{a}$}{telD}
			
			\ed{u4}{v4}{red}
			\dted{u4}{vv4}{red}
			\ed{uu4}{v4}{red}
			\ded{uu4}{vv4}{lightgray}
			
			\vcl{(4,2)}{u5}{}{white}
			\vcl{(4,3)}{v5}{}{white}
			\tvcl{(5,3)}{uu5}{$\B{b}$}{telD}
			\vcl{(5,2)}{vv5}{}{white}
			
			\ed{u5}{v5}{red}
			\ed{u5}{vv5}{red}
			\dted{uu5}{v5}{red}
			\ded{uu5}{vv5}{lightgray}
			
			\ed{u4}{vv2}{black}
			\ed{v4}{vv3}{black}
			\ed{u5}{uu3}{black}			
			
			\ed{v5}{uu1}{black}
			\ed{vv5}{uu4}{black}
			
			\tvcl{(-1,0)}{t1}{$\B{A}$}{telS}
			\tvcl{(-1,3)}{t2}{$\B{B}$}{telS}
			
			\dted{t1}{u2}{black}
			\dted{t2}{v1}{black}
			
			\node at (2.5,3.3) {$\beta$};
			\node at (2.5,-0.3) {$\alpha$};
			
		\end{tikzpicture}
	}
	
	\medskip
	\scalebox{0.9}{
		\begin{tikzpicture}[scale=0.9]
			
			\pvcl{(1,1.2)}{p11}{\scriptsize $\B{B}\beta\B{b}$}{yellow}
			\pvcl{(1,-3.2)}{p22}{\scriptsize $\B{A}\alpha\B{a}$}{yellow}
			
			\cvcl{(0,-1)}{c1}{$\mathcal{Y}_1$}{yellow}
			\cvcl{(1,0)}{c4}{$\mathcal{Y}_4$}{mypink}
			\cvcl{(1,-2)}{c2}{$\mathcal{Y}_2$}{mypink}
			\cvcl{(2,-1)}{c3}{$\mathcal{Y}_3$}{yellow}
			\ed{c1}{c2}{black}
			\ed{c2}{c3}{black}
			\ed{c3}{c4}{black}
			\ed{c4}{c1}{black}
			\ed{c2}{p22}{black}
			\ed{c4}{p11}{black}

		\end{tikzpicture}
    }
    
	\end{center}
	\end{minipage}
	\hspace{5mm}
	\begin{minipage}{5cm} 
    \begin{center}
     \textbf{(iii)}
    \smallskip
    
    \scalebox{0.8}{
		\begin{tikzpicture}[scale=0.8]
			
			\vcl{(0,2)}{u1}{}{white}
			\vcl{(0,3)}{v1}{}{white}
			\vcl{(1,3)}{uu1}{}{white}
			\vcl{(1,2)}{vv1}{}{white}

			\node at (0.5,2.5) {\color{darkred}\scriptsize 1};
			\node at (0.5,0.5) {\color{darkred}\scriptsize 2};
			\node at (2.5,1.5) {\color{darkred}\scriptsize 3};
			\node at (4.5,0.5) {\color{darkred}\scriptsize 4};
			\node at (4.5,2.5) {\color{darkred}\scriptsize 5};
			
			\node at (1,1.5) {$\mathcal{Y}_1$};
			\node at (2.5,0.5) {$\mathcal{Y}_2$};
			\node at (4,1.5) {$\mathcal{Y}_3$};
			\node at (2.5,2.5) {$\mathcal{Y}_4$};
						
			\dted{u1}{v1}{red}
			\ed{u1}{vv1}{red}
			\dted{uu1}{v1}{red}
			\ed{uu1}{vv1}{red}
			
			\node at (-0.3,1.5) {$\alpha$};
			
			\tvcl{(0,0)}{u2}{$\B{a}$}{telD}
			\vcl{(0,1)}{v2}{}{white}
			\vcl{(1,1)}{uu2}{}{white}
			\vcl{(1,0)}{vv2}{}{white}
			
			\dted{u2}{v2}{red}
			\ded{u2}{vv2}{lightgray}
			\ed{uu2}{v2}{red}
			\ed{uu2}{vv2}{red}
			
			\vcl{(2,1)}{u3}{}{white}
			\vcl{(2,2)}{v3}{}{white}
			\vcl{(3,2)}{uu3}{}{white}
			\vcl{(3,1)}{vv3}{}{white}
			
			\ed{u3}{v3}{red}
			\ed{u3}{vv3}{red}
			\ed{uu3}{v3}{red}
			\ed{uu3}{vv3}{red}
			
			\ed{vv1}{v3}{black}
			\ed{u3}{uu2}{black}
			\ed{u1}{v2}{black}
			
			\vcl{(4,0)}{u4}{}{white}
			\vcl{(4,1)}{v4}{}{white}
			\vcl{(5,1)}{uu4}{}{white}
			\vcl{(5,0)}{vv4}{}{white}
			
			\ed{u4}{v4}{red}
			\ded{u4}{vv4}{lightgray}
			\ed{uu4}{v4}{red}
			\ded{uu4}{vv4}{lightgray}

			\node at (2.5,3.3) {$\beta$};
			
			\vcl{(4,2)}{u5}{}{white}
			\vcl{(4,3)}{v5}{}{white}
			\tvcl{(5,3)}{uu5}{$\B{b}$}{telD}
			\vcl{(5,2)}{vv5}{}{white}
			
			\ed{u5}{v5}{red}
			\ed{u5}{vv5}{red}
			\dted{uu5}{v5}{red}
			\ded{uu5}{vv5}{lightgray}
			
			\ed{u4}{vv2}{black}
			\ed{v4}{vv3}{black}
			\ed{u5}{uu3}{black}			
			
			\ed{v5}{uu1}{black}
			\ed{vv5}{uu4}{black}
			
			\tvcl{(-1,3)}{t1}{$\B{A}$}{telS}
			
			\dted{t1}{v1}{black}

		\end{tikzpicture}
	}
	
	\medskip
	\scalebox{0.9}{
		\begin{tikzpicture}[scale=0.9]
			
			\pvcl{(-1.2,1.5)}{p11}{\scriptsize $\B{A}\beta\B{b}$}{yellow}
			\pvcl{(-1.2,0)}{p22}{\scriptsize $\B{A}\alpha\B{a}$}{mylime}
			
			\cvcl{(0,0)}{c1}{$\mathcal{Y}_1$}{yellow}
			\cvcl{(0,1.5)}{c4}{$\mathcal{Y}_4$}{mylime}
			\cvcl{(1.5,0)}{c2}{$\mathcal{Y}_2$}{mylime}
			\cvcl{(1.5,1.5)}{c3}{$\mathcal{Y}_3$}{yellow}
			\ed{c1}{c2}{black}
			\ed{c2}{c3}{black}
			\ed{c3}{c4}{black}
			\ed{c4}{c1}{black}
			\ed{c1}{p22}{black}
			\ed{c4}{p11}{black}
			\ed{p22}{p11}{black}

		\end{tikzpicture}
    }
    
	\end{center}
	\end{minipage}

	\end{center}

	\caption{All possibilities for a bubble of four 6-cycles with intersecting paths. (i) is symmetrically surrounded by a cyclic path-line of length 4. (ii) has has two paths and two connections between these paths and cycles from the same independent set. (iii) has two paths forming a path line and two connections between these paths and cycles from distinct independent sets.}\label{afig:int4bub-sat}

\end{figure}

\smallskip
In Figures~\ref{afig:int2bub-sat}-\ref{afig:int4bub-sat}  we enumerate all cases of balanced cycle-bubbles that have at most 8 cycles and intersect with more than one path. We omit the general and well described case of a cycle-line with plug connections.
In all figures, dotted red edges are exclusively for paths, dashed gray edges are pruned out, blue nodes represent $\g{S}$-telomeres and gray nodes represent $\check{\g{D}}$-telomeres. Furthermore, green/yellow solutions are co-optimal, while yellow solutions are better than the pink alternatives.

\smallskip

\input{appendix/Supplementary-Figures/bubble2}

\refstepcounter{myfigure}


\end{document}